\documentclass[aps,prb,twocolumn,superscriptaddress,showpacs,amsmath,amssymb]{revtex4-1}

\usepackage{amssymb,amsbsy,amsmath}

\usepackage{graphicx}
\usepackage{dcolumn}
\usepackage{bm}
\usepackage{subfigure}
\usepackage{color}

\newcommand{\op}[1]{%
    \fontdimen12\textfont3=2pt\fontdimen12\scriptfont3=1.4pt%
    \!\null\mathop{\vphantom{#1}\smash{#1}}\limits_{\sim}\null\!}
\newcommand{\xref}[1]{\protect\ref{#1}}
\newcommand{\figref}[1]{Fig.~\protect\ref{#1}}
\newcommand{\fmref}[1]{(\protect\ref{#1})}

\def\ket#1{\, | \, {#1} \, \rangle}
\newcommand{\braket}[2]{\langle \, {#1} \, | \, {#2} \, \rangle}

\newcommand{\MnAC}{\protect Mn$_{12}$-acetate}
\begin{document}
\title{Thermodynamic observables of Mn$_{12}$-acetate calculated
for the full spin-Hamiltonian}

\author{Oliver Hanebaum}
\author{J\"urgen Schnack}
\email{jschnack@uni-bielefeld.de}
\affiliation{Fakult\"at f\"ur Physik, Universit\"at Bielefeld, Postfach 100131, D-33501 Bielefeld, Germany}

\date{\today}

\begin{abstract}
35 years after its synthesis magnetic observables are calculated
for the first time for the molecular nanomagnet
Mn$_{12}$-acetate using a spin-Hamiltonian that contains all
spins. Starting from a very advanced DFT
parameterization [Phys. Rev. B 89, 214422] we evaluate
magnetization and specific heat for 
this anisotropic system of 12 manganese ions with a staggering
Hilbert space dimension of 100,000,000 using the
Finite-Temperature Lanczos Method. 
\end{abstract}


\maketitle

\section{Introduction}
\label{sec-1}

Density Functional Theory (DFT) has greatly advanced over the past years 
and is nowadays able to predict parameters of spin-Hamiltonians
with which the low-temperature physics of correlated magnetic 
materials can be described, compare e.g. 
Refs.~\onlinecite{LKA:JMMM87,RAA:JACS97,KHP:PRL01,RHD:JMMM02,BKM:PRB03,BKP:PRB2004,ZMS:PRB08A,ZPS:IC13,KDT:DT13,CCS:PRL13,SFF:PRB14,PLM:ACIE14,SFR:CAEJ14,SiR:CAEJ14,KGD:PRB15}.
Along this line the complex spin-Hamiltonian of one of the most
exciting magnetic molecules, \MnAC, was recently predicted.\cite{MKJ:PRB14}
These calculations consider almost all terms that are bilinear in
spin operators as there are the Heisenberg exchange interaction,
the anisotropic antisymmetric exchange interaction and the
single-ion anisotropy tensors. The new calculations outperform
earlier attempts\cite{KBB:PRB02,BLD:PRB02} and provide much richer 
electronic insight 
than parameterizations obtained from fits to magnetic 
observables\cite{CSO:PRB04} or parameterizations resting on 
knowledge from similar but smaller systems. But despite all the
success DFT is not capable of evaluating 
magnetic observables which is the reason for the detour via
spin-Hamiltonians. 

The magnetism of anisotropic molecular spin systems is
fascinating due to interesting phenomena such as 
bistability and quantum tunneling of the
magnetization.\cite{GSV:2006} Bistability in connection
with a small tunneling rate leads to a magnetic hysteresis 
of molecular origin in these systems. That's why such 
molecules are termed Single
Molecule Magnets (SMM);  Mn$_{12}$-acetate is the most prominent SMM.
\cite{Lis:ACB80,STS:JACS93,SGC:Nat93,TLB:Nature96,GNS:PRB98,CAG:JMMM01,GaS:ACIE03} 
But although \MnAC\ contains only four Mn$^{\text{(IV)}}$ ions
with $s=3/2$ and eight Mn$^{\text{(III)}}$ ions with 
$s=2$, it
constitutes a massive challenge for theoretical calculations in
terms of spin-Hamiltonians since the underlying Hilbert space 
of dimension 100,000,000
is orders of magnitude too big for an exact and complete
matrix diagonalization.\cite{ScS:PRB09} But thanks to the fact
that the 
zero-field split ground-state multiplet is energetically
separated from higher-lying levels a description using only the
$S=10$ ground-state manifold is sufficient to explain
observables at low temperature -- this approach was used in the
past. Thermodynamic functions which involve
higher-lying levels, for instance observables at higher
temperature, can of course not be evaluated in such an
approximation. 

Fortunately not only on the side of DFT progress has been made
in past years, but also in terms of powerful approximations 
for spin-Hamiltonian calculations. For not too big systems with
Hilbert spaces with dimensions of up to $10^{10}$ Krylov space methods
such as the Finite-Temperature Lanczos Method
(FTLM) have proven to provide astonishingly accurate
approximations of magnetic 
observables.\cite{PhysRevB.49.5065,JaP:AP00,MHL:CPL01,LPE:PRB03,ADE:PRB03,WWA:RMP06,PrB:SSSSS13,SSP:EPJB04,ZST:PRB06,STS:PRB07,SSZ:NJP12,ScW:EPJB10,ScH:EPJB13}
While FTLM has been used for Heisenberg spin systems mostly, very
recently the method was advanced to anisotropic spin systems.\cite{HaS:EPJB14}

In this article we therefore employ the most recent FTLM
in order to study thermodynamic functions of \MnAC\ starting
from parameterizations provided by DFT or other methods.
We evaluate the magnetization as well as the specific heat 
both as function of temperature and
field and compare the various parameterizations of the
spin-Hamiltonian. 

The article is organized as follows. In Section~\xref{sec-2} the
employed Hamiltonian as well as 
basics of the Finite-Temperature Lanczos Method are introduced. 
Section~\xref{sec-3}, \xref{sec-4} and \xref{sec-5} discuss the
effective magnetic moment, the magnetization and the specific
heat, respectively. The article closes with summary and outlook.

\section{FTLM for anisotropic spin systems}
\label{sec-2}

For \MnAC, which is a highly anisotropic spin system, 
the complete Hamiltonian of the spin system is given by the
exchange term, the single-ion anisotropy, and the Zeeman term, 
i. e.  
\begin{eqnarray}
\label{E-2-1}
\op{H}
&=&
\sum_{i<j}\;
\op{\vec{s}}_i \cdot {\mathbf J}_{ij} \cdot \op{\vec{s}}_j
+
\sum_{i}\;
\op{\vec{s}}_i \cdot 
{\mathbf D}_i
\cdot \op{\vec{s}}_i
\\
&&+
\mu_B\, B\,
\sum_{i}\;
g_i
\op{s}^z_i
\nonumber
\ .
\end{eqnarray}
${\mathbf J}_{ij}$ is a $3\times 3$ matrix for each interacting
pair of spins at sites $i$ and $j$ which contains the
isotropic Heisenberg exchange parameters together with the
anisotropic symmetric and antisymmetric terms.
In the sign convention of \fmref{E-2-1} a positive
Heisenberg exchange corresponds to an
antiferromagnetic interaction and a negative one to a ferromagnetic
interaction. 
${\mathbf D}_i$ denotes the single-ion anisotropy tensor at site
$i$, which in its eigensystem $\vec{e}_{i}^{\,1}$, $\vec{e}_{i}^{\,2}$,
$\vec{e}_{i}^{\,3}$, can be decomposed as
\begin{eqnarray}
\label{E-2-2}
{\mathbf D}_i
&=&
D_i \vec{e}_{i}^{\,3} \otimes \vec{e}_{i}^{\,3}
+
E_i
 \left\{
\vec{e}_{i}^{\,1} \otimes \vec{e}_{i}^{\,1}
- 
\vec{e}_{i}^{\,2} \otimes \vec{e}_{i}^{\,2}
\right\}\ .
\end{eqnarray}
The terms $g_i$ could in general be $3\times 3$ matrices, too,
but for the sake of simplicity it is assumed that the $g_i$ are
numbers and moreover that $g_i=2$ for all ions. This assumption is justified
for the Mn$^{\text{(IV)}}$ and Mn$^{\text{(III)}}$ ions in
\MnAC, since the 
$g$-factors of both ions are estimated to be very close to
two.\cite{BGS:PRB97,CSO:PRB04,GHK:IC09,GHT:DT10,HGM:CS12} 

The Finite-Temperature Lanczos Method (FTLM) approximates the 
partition function in two ways:\cite{PhysRevB.49.5065,JaP:AP00} 
\begin{eqnarray}
\label{E-2-3}
Z(T,B)
&\approx&
\frac{\text{dim}({\mathcal H})}{R}
\sum_{\nu=1}^R\;
\sum_{n=1}^{N_L}\;
e^{-\beta \epsilon_n^{(\nu)}} |\braket{n(\nu)}{\nu}|^2
\ .
\end{eqnarray}
The sum over a complete set of vectors
is replaced by a much smaller sum over $R$ random vectors $\ket{\nu}$.
The exponential of the Hamiltonian is then approximated by
its spectral representation in a Krylov space spanned by the
$N_L$ Lanczos vectors starting from the respective random vector
$\ket{\nu}$. $\ket{n(\nu)}$ is the n-th eigenvector of $\op{H}$ in
this Krylov space. 

It turns out that a very good accuracy can already
be achieved for parameters $R\approx 10$ and $N_L\approx 100$,
especially in cases when the low-lying energy spectrum is
dense.\cite{ScW:EPJB10,HaS:EPJB14}

\section{Effective magnetic moment as function of temperature}
\label{sec-3}

\begin{figure}[ht!]
\centering
\includegraphics*[clip,width=60mm]{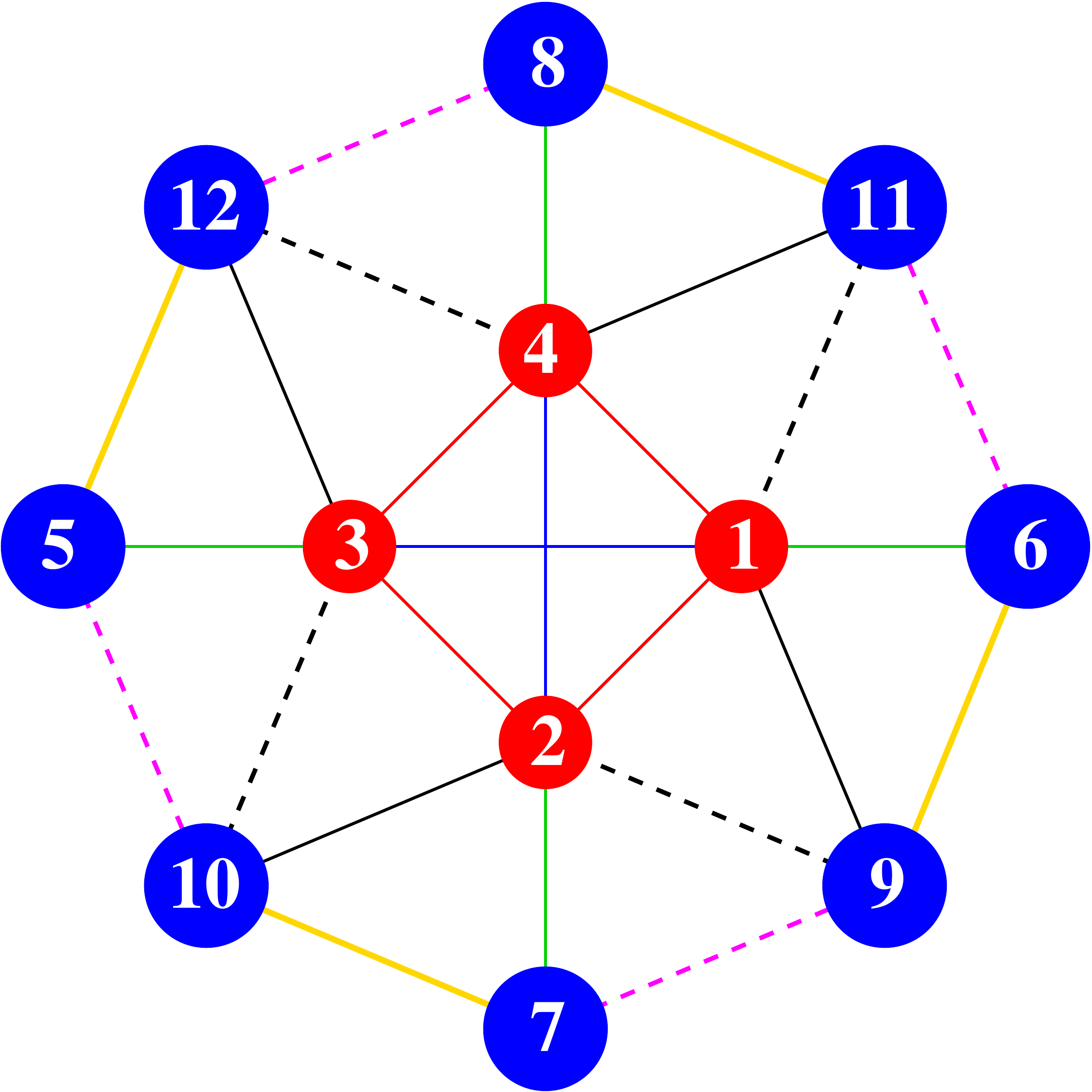}
\caption{(Color online) Schematic structure of Mn12, same labeling as in 
  Ref.~\onlinecite{MKJ:PRB14}. Mn$^{\text{(IV)}}$ ions (1-4) are shown as red circles, 
Mn$^{\text{(III)}}$ ions (5-12) as blue ones. An $S_4$ symmetry is assumed.}   
\label{tlmm-f-1}
\end{figure}

\MnAC\ contains four Mn$^{\text{(IV)}}$ ions
with $s=3/2$ and eight Mn$^{\text{(III)}}$ ions with 
$s=2$. Following Ref.~\onlinecite{MKJ:PRB14} the ions and the
exchange pathways are depicted in \figref{tlmm-f-1}. 
Mn$^{\text{(IV)}}$ ions (1-4) are shown as red circles, 
Mn$^{\text{(III)}}$ ions (5-12) as blue ones.
An $S_4$ symmetry of the molecule is assumed.\cite{FCP:CEC13}

Since the discovery of the pronounced SMM-properties of
\MnAC\ several groups developed parameterizations of the full
spin-Hamiltonian. These data sets, of which the most prominent
ones are given in Table~\xref{tab-j}, contain parameterizations
of Heisenberg models and were put forward
following various scientific reasonings. Earlier attempts
assigned values of exchange interactions in analogy to smaller
compounds with similar chemical bridges between the manganese
ions. Later investigations combined for instance
high-temperature series expansion with the evaluation of
low-lying excitations seen in Inelastic Neutron Scattering (INS)
experiments.\cite{CSO:PRB04} 
A necessary condition that has to
be met by all parameterizations, is that the ground state has a
total spin of $S=10$. The most recent DFT parameterization is
also compatible with INS experiments.\cite{MKJ:PRB14}

\begin{table}[!h]
\centering
\caption{Intra-molecular isotropic exchange interaction
  parameters (in meV) as suggested by various authors, compare
  Ref.~\onlinecite{MKJ:PRB14}. 
  The spin labels are explained in \figref{tlmm-f-1}.}
\label{tab-j}
\begin{tabular}{lllllcccc}
  \hline
  \hline
No.\phantom{.} &  Bond (i-j) \phantom{.} & 1-6 \phantom{.} & 1-11 \phantom{.} & 1-9 \phantom{.} & 6-9 \phantom{.} & 7-9 \phantom{.}  & 1-4 \phantom{.} & \phantom{a} 1-3 \\
  \hline
1 &J$_{ij}$ (Ref.~\onlinecite{MKJ:PRB14})   & 4.6  & 1.0   & 1.7  & -0.45 & -0.37 & -1.55 & -0.5 \\
2 &J$_{ij}$ (Ref.~\onlinecite{BLD:PRB02})   & 4.8  & 1.37  & 1.37 & -0.5  & -0.5 & -1.6 & -0.7 \\
3 &J$_{ij}$ (Ref.~\onlinecite{CSO:PRB04})   & 5.8  & 5.3   & 5.3  &   0.5  &  0.5 & 0.7 & 0.7 \\
4 &J$_{ij}$ (Ref.~\onlinecite{BGM:1997})    & 7.4  & 1.72  & 1.72 & 0 & 0 & -1.98 & 0 \\
5 &J$_{ij}$ (Ref.~\onlinecite{RJS:PRB02})   & 10.25& 10.17 & 10.17& 1.98 & 1.98 & -0.69 & -0.69 \\
\hline
\end {tabular}
\end {table}

\begin{figure}[ht!]
\centering
\includegraphics*[clip,width=70mm]{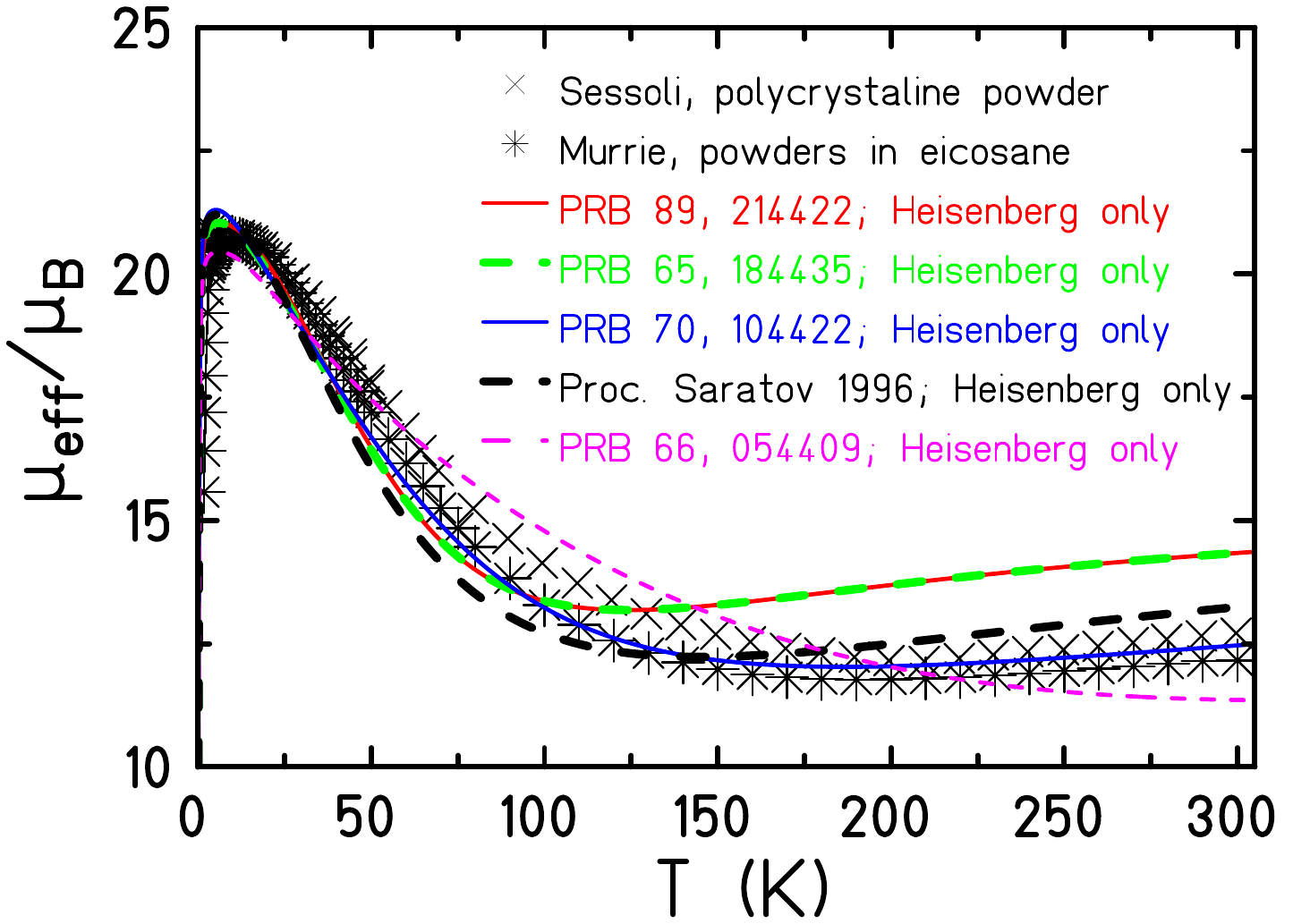}

\includegraphics*[clip,width=70mm]{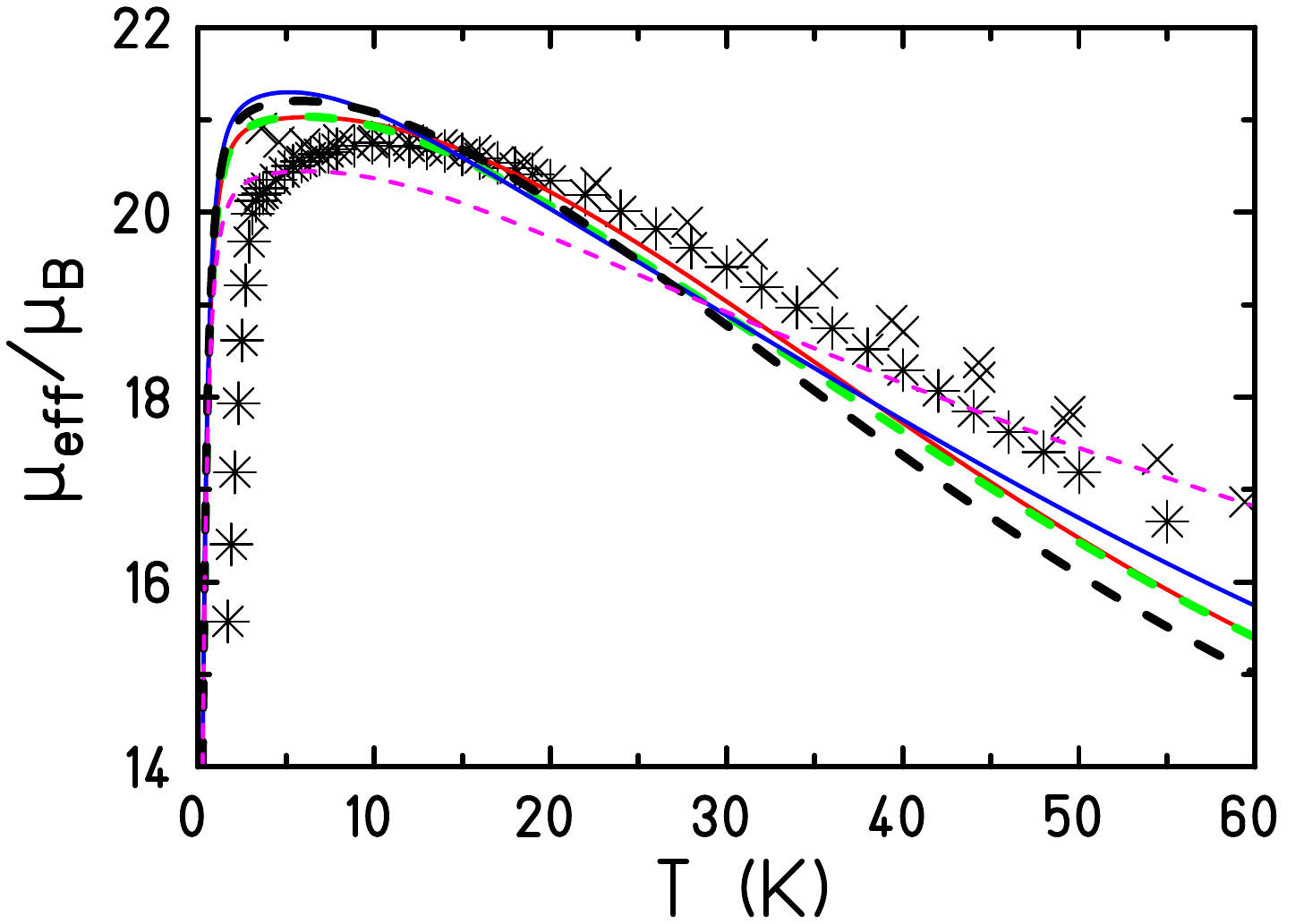}
\caption{(Color online) Effective magnetic moment of \MnAC\ at
  $B=0.1$~T. Data of Sessoli\cite{GaS:ACIE03} and
  Murrie\cite{Par:PhD10} are given by symbols. Observables employing
  the Heisenberg part of parameterizations only are displayed by
  curves. The parameterizations correspond to those given in
  Table~\xref{tab-j} from top to bottom.}    
\label{tlmm-f-b}
\end{figure}

Figure \xref{tlmm-f-b} shows the effective magnetic moment at a small
external field of $B=0.1$~T as a function of temperature.
Data of Sessoli\cite{GaS:ACIE03} and Murrie\cite{Par:PhD10} are
given by symbols. For the theory curves only the Heisenberg part
of the respective 
parametrizations is used. Since the
Heisenberg model is SU(2) symmetric, a FTLM version employing
total $\op{S}^z$ symmetry was used with $R=100$ and
$N_L=120$ in this case.\cite{ScW:EPJB10} One realizes that the gross
structure of the magnetic moment (which is proportional to
$\sqrt{\chi T}$) is achieved by all parameterizations especially
at lower temperatures of $T\lessapprox 50$~K. A finer inspection
shows that the maximum is at a too low temperature for all
parameterizations, so that the experimental low-temperature data
points are 
not met. For higher temperatures towards room temperature one
notices that only one parameterization\cite{CSO:PRB04} (blue
curve) closely follows the experimental data towards the
paramagnetic limit. This 
is not astonishing since this parameterization was fitted to the
high-temperature tail using a high-temperature series
expansion. We
conjecture that the somewhat too large effective magnetic moment of the
DFT parameterizations, Refs.~\onlinecite{MKJ:PRB14,BLD:PRB02}, at room
temperature are related to the fact, that these
parameterizations contain ferromagnetic interactions whereas a
fit to the high-temperature behavior,
Ref.~\onlinecite{CSO:PRB04}, leads only to antiferromagnetic
interactions, compare Table~\xref{tab-j}. In addition the
antiferromagnetic interactions 1-11 and 1-9 are much stronger in
Ref.~\onlinecite{CSO:PRB04}. 

\begin{figure}[ht!]
\centering
\includegraphics*[clip,width=70mm]{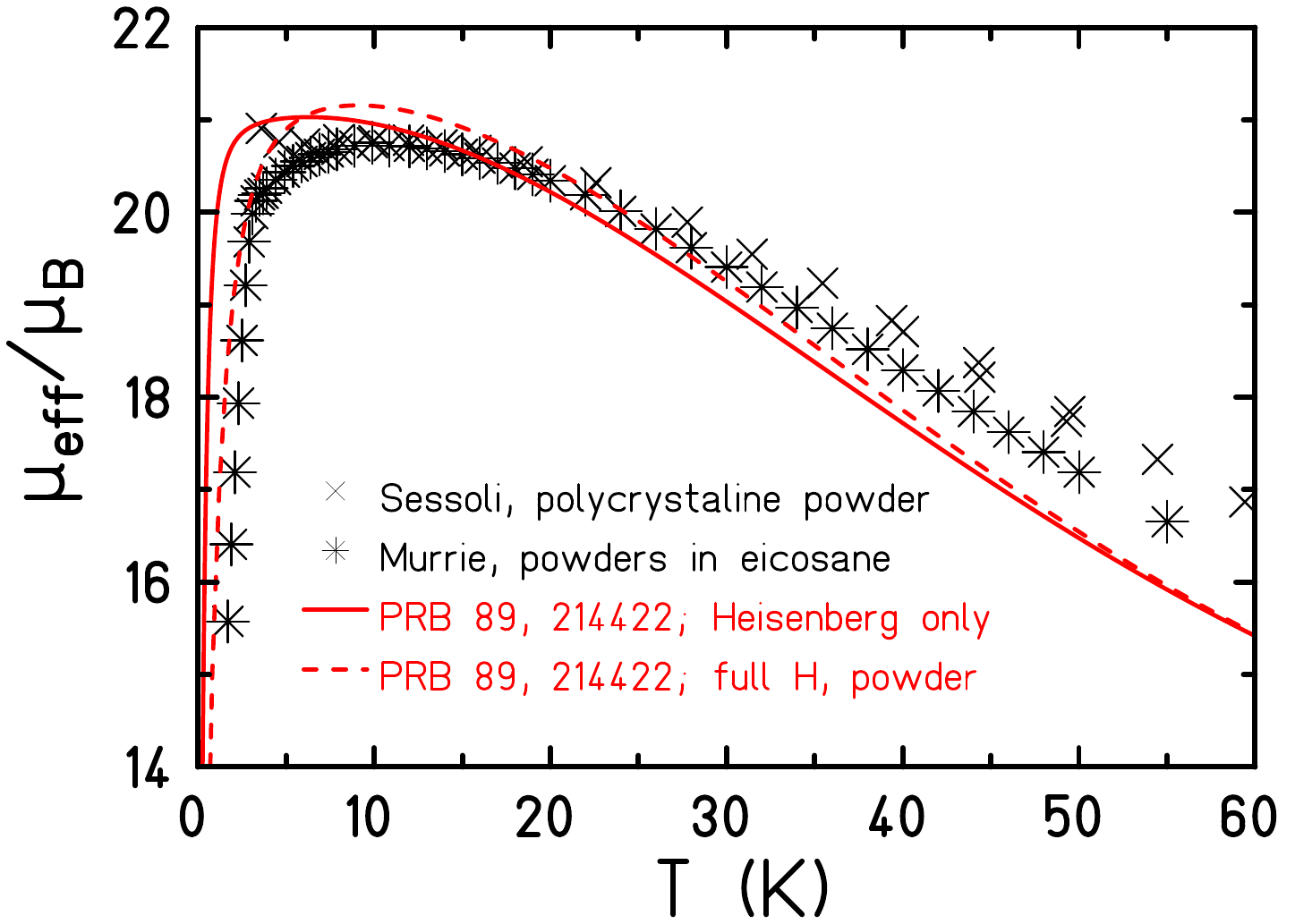}
\caption{(Color online) Effective magnetic moment of \MnAC\ at
  $B=0.1$~T, same as \figref{tlmm-f-b}. The dashed curve shows
  the result of a calculation employing the full Hamiltonian of
  Ref.~\onlinecite{MKJ:PRB14}. The powder average is performed
  over a regular grid of 20 directions on the unit
  sphere.\cite{Sch:CMP09}}    
\label{tlmm-f-f}
\end{figure}

Using the recently developed FTLM for anisotropic
systems\cite{HaS:EPJB14} we could calculate the effective
magnetic moment starting from the DFT parameterization of
Ref.~\onlinecite{MKJ:PRB14}. 
Besides the Heisenberg terms of Table~\xref{tab-j} 
this parameterization contains anisotropic Dzyaloshinskii-Moriya
interactions as well as full $3\times 3$ anisotropy tensors for
each manganese ion. It turns out that the additional terms
improve the low-temperature data, compare \figref{tlmm-f-f}. The
maximum shifts towards the experimental position and the data
points for smaller temperatures are much better
approximated. None of the plain Heisenberg models could achieve
such an improvement so far. For temperatures above 60~K the
anisotropic terms are irrelevant.

\section{Magnetization as function of applied field}
\label{sec-4}

The low-temperature magnetization usually provides strong
fingerprints of the underlying spin-Hamiltonian for instance in
the case of magnetization steps due to ground state level
crossings. For \MnAC\ the magnetization shows even richer
characteristics, since below the blocking temperature a magnetic
hysteresis is observed.\cite{SGC:Nat93} This exciting physical
property turns out to constitute a problem, when comparing to
the theoretical equilibrium magnetization. Due to the long
relaxation times, approximately 2800 hours at $T=2$~K,\cite{GSV:2006}
the experimental values do not necessarily 
reflect equilibrium values. On the other hand, the theoretical
evaluation of 
non-equilibrium observables for a full spin model of \MnAC\ is
totally out of reach.

\begin{figure}[ht!]
\centering
\includegraphics*[clip,width=70mm]{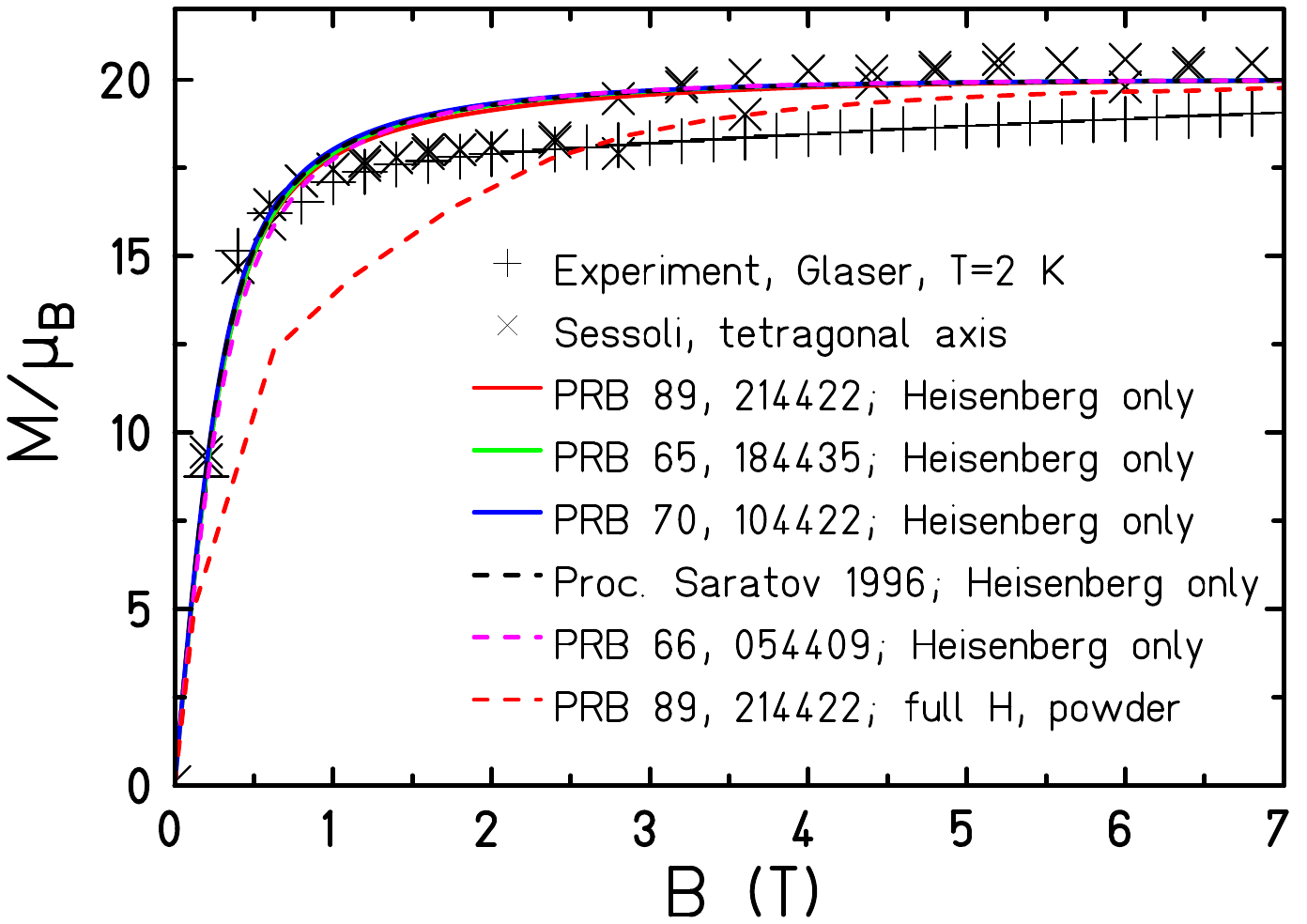}
\caption{(Color online) Magnetization of \MnAC\ at
  $T=2$~K. Color code of curves as above. The dashed curve is
  evaluated for two field values per 1 T field interval only.}   
\label{tlmm-f-c}
\end{figure}

Figure~\xref{tlmm-f-c} provides two experimental data sets
as well as various theoretical curves. The data set of
Glaser\cite{GlW:PC15} was taken on a powder sample whereas the
data set of Sessoli\cite{GaS:ACIE03} was taken on a single
crystal with a field in
direction of the tetragonal axis of the $S_4$ symmetric
molecule. Both data sets coincide up to $B\approx 2.5$~T, then
the magnetization along the tetragonal axis jumps whereas the
powder signal smoothly increases with field. Already at this
point it becomes clear that the measurements cannot reflect
equilibrium properties, because the magnetization along the
tetragonal axis, which is the easy axis of this strongly
anisotropic molecule,\cite{NSC:JMMM95} cannot be the same as the
powder averaged magnetization. 

Interestingly, all theory curves that rest on Heisenberg model
calculations agree with each other perfectly, which is due to 
the fact that all produce a $S=10$ ground state that is largely
separated from excited levels. They also agree with
the experimental magnetization up to a field of $B\approx
1$~T. Between 1~T and 2.7~T the experimental data points stay below the theoretical
curves. Above $B\approx 2.7$~T theory and magnetization along
the tetragonal axis meet again. The calculation for the full
spin model of  \MnAC\ as given in Ref.~\onlinecite{MKJ:PRB14}
yields a rather unexpected result: the powder-averaged
magnetization stays well below all other theory curves (expected
since anisotropic), but stays also well below both experimental
curves (unexpected at least compared to the experimental powder
data). 

\begin{figure}[ht!]
\centering
\includegraphics*[clip,width=70mm]{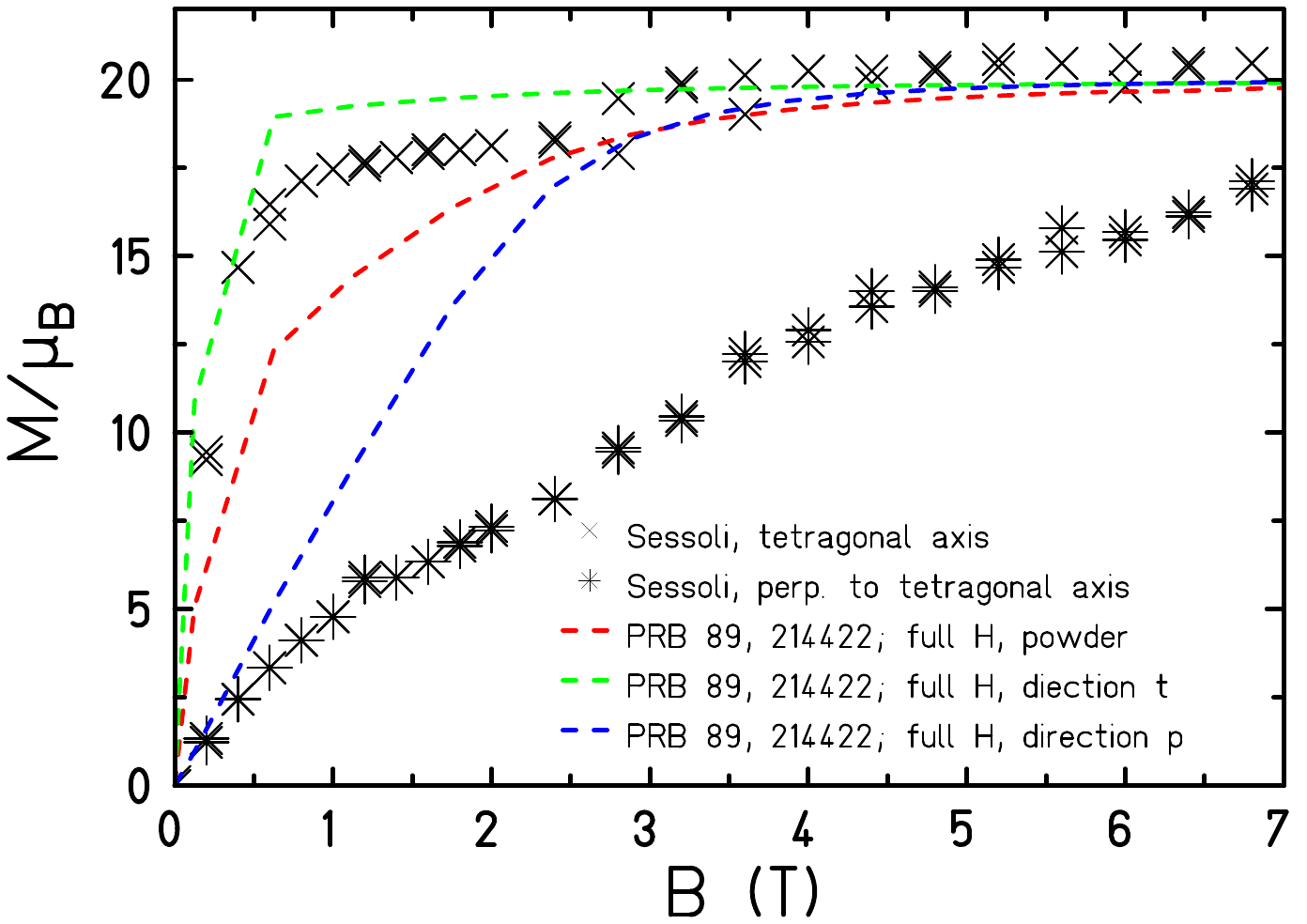}
\caption{(Color online) Magnetization of \MnAC\ at $T=2$~K. 
  The dashed curves are
  evaluated for two field values per 1 T field interval
  only. Direction $t=(0.0, -0.35682, 0.93417)$, 
  direction $p=(-0.35682, 0.93417, 0.0)$.}   
\label{tlmm-f-i}
\end{figure}

In the following we compare our results with the
measurements of Ref.~\onlinecite{GaS:ACIE03} along two different
directions, along the tetragonal axis and perpendicular to that,
i.e. somewhere in the $xy$-plane. Figure~\xref{tlmm-f-i}
presents three theory curves: one for the powder average and two
along special directions. Direction $t=(0.0, -0.35682, 0.93417)$
points roughly along the tetragonal axis (inclination of about
$20^\circ$) and direction 
$p=(-0.35682, 0.93417, 0.0)$ lays in the $xy$-plane. Both
theoretical curves show systematically larger magnetization
values than the experiment. This could be for two reasons:
either the experimental curves are not in equilibrium, which is
possible at $T=2$~K where the relaxation time of \MnAC\ is long
or the parameterization of Ref.~\onlinecite{MKJ:PRB14} is still
not yet optimal, i.e. the ${\mathbf D}$ tensors could be too
weak, for instance. Nevertheless, the positive message is, that we
can now calculate such curves and compare with experimental
data. The theoretical costs, by the way, are still enormous. For
the investigations shown in this article about 2~Mio. CPU hours
on a supercomputer had to be used, since the FTLM procedure has
to be completed twice for every field value and direction. Therefore,
only a few field values have been used for the theoretical magnetization curves.

\begin{figure}[ht!]
\centering
\includegraphics*[clip,width=70mm]{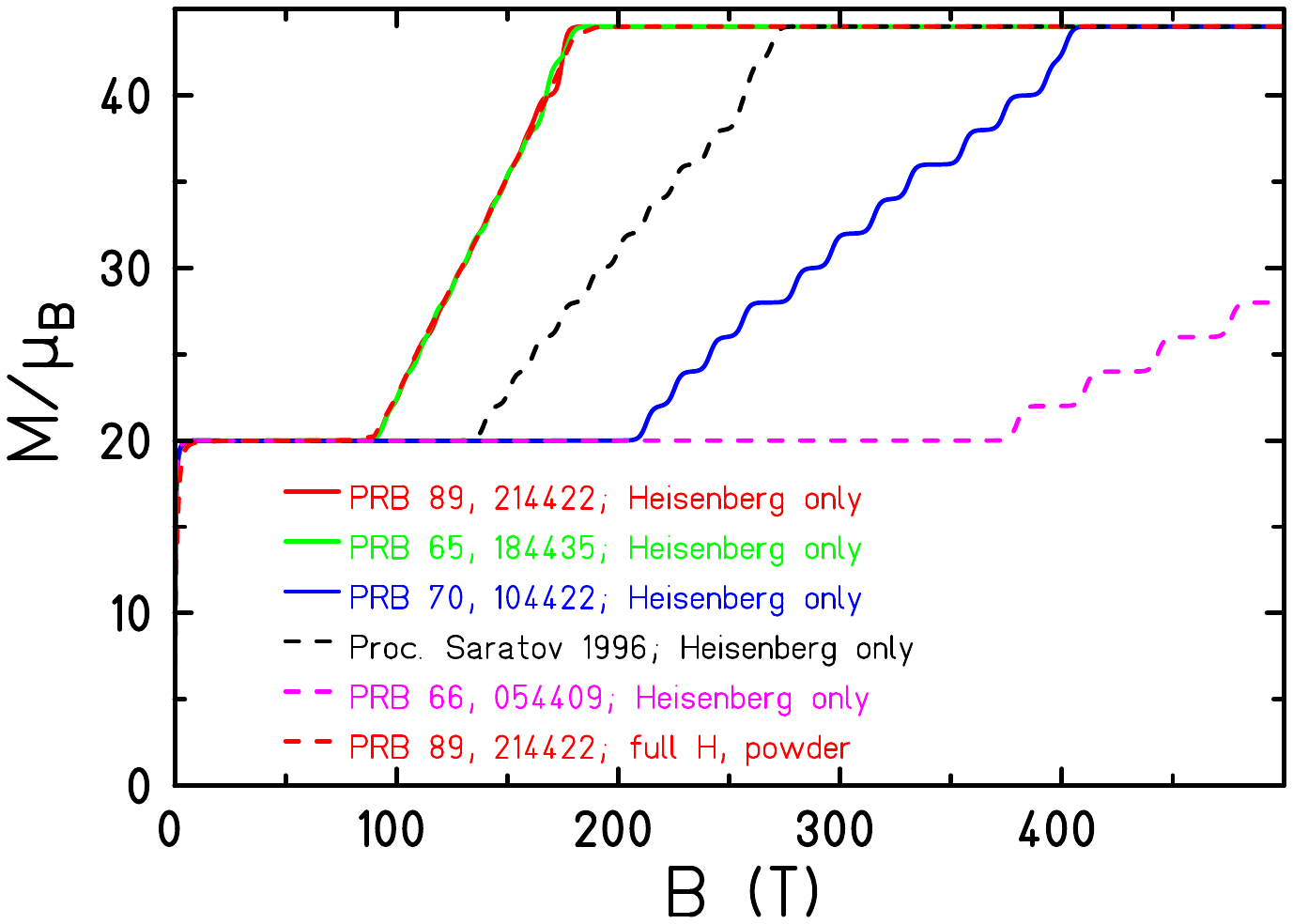}
\caption{(Color online) High-field magnetization of \MnAC\ at $T=2$~K.}   
\label{tlmm-f-h}
\end{figure}

Finally we would like to present the high-field magnetization
curves. As can be seen in \figref{tlmm-f-h} the
various parameterizations lead to distinctive differences at
high fields. The high-field magnetization could be and has been
measured in megagauss experiments.\cite{ZLL:PU98} Interestingly, the
magnetization data given in Ref.~\onlinecite{ZLL:PU98} show
pronounced features, likely related to magnetization steps,
between 180~T and 400~T which could be compatible with the
parameterization of Ref.~\onlinecite{CSO:PRB04} (blue curve in
\figref{tlmm-f-h}). As realized already by the authors, this
parameterization produces a sequence of 
level crossings between the $S=10$ ground manifold and the fully
polarized state exactly in this field range.

\section{Heat capacity}
\label{sec-5}

\begin{figure}[ht!]
\centering
\includegraphics*[clip,width=70mm]{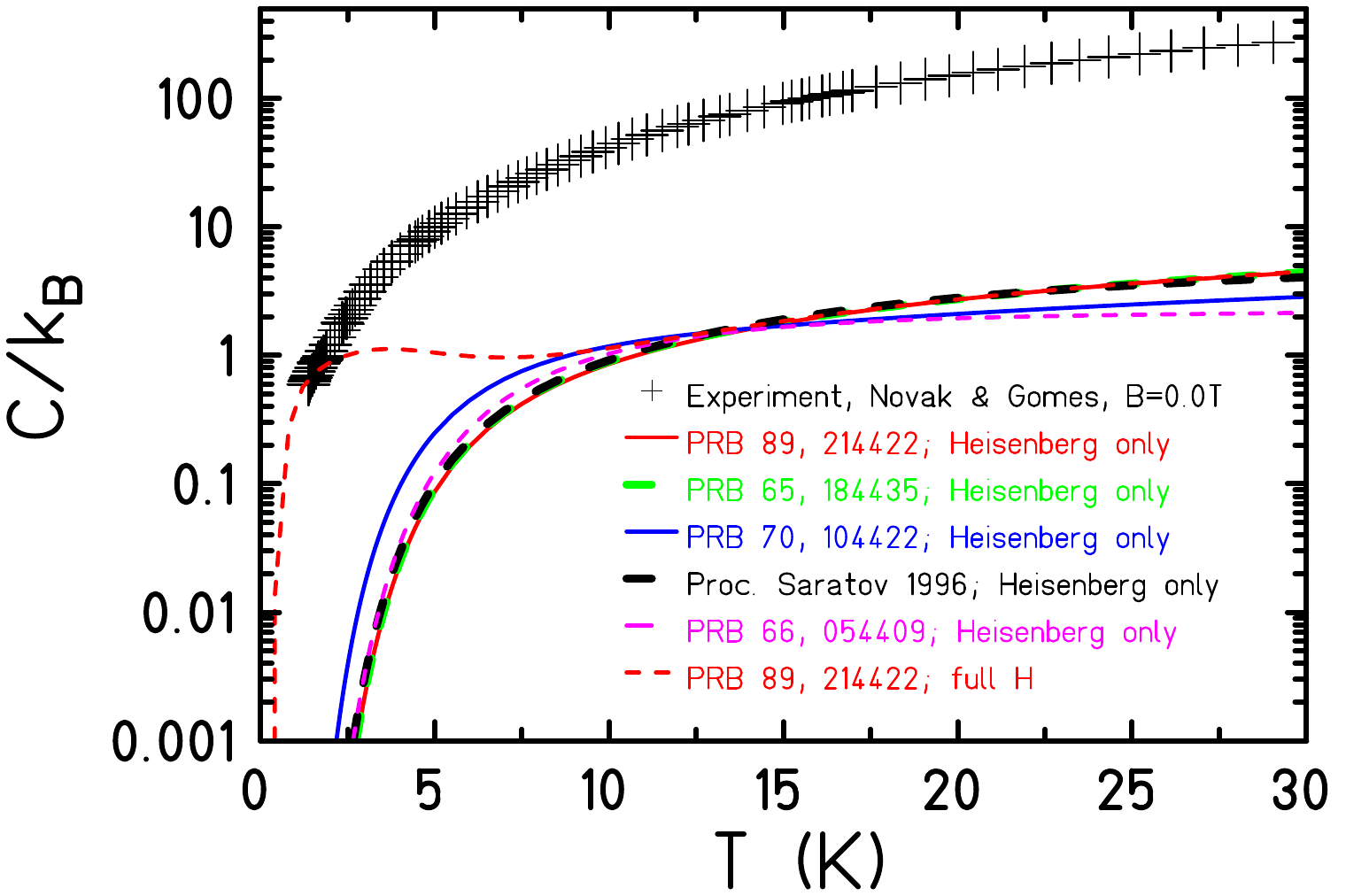}

\includegraphics*[clip,width=70mm]{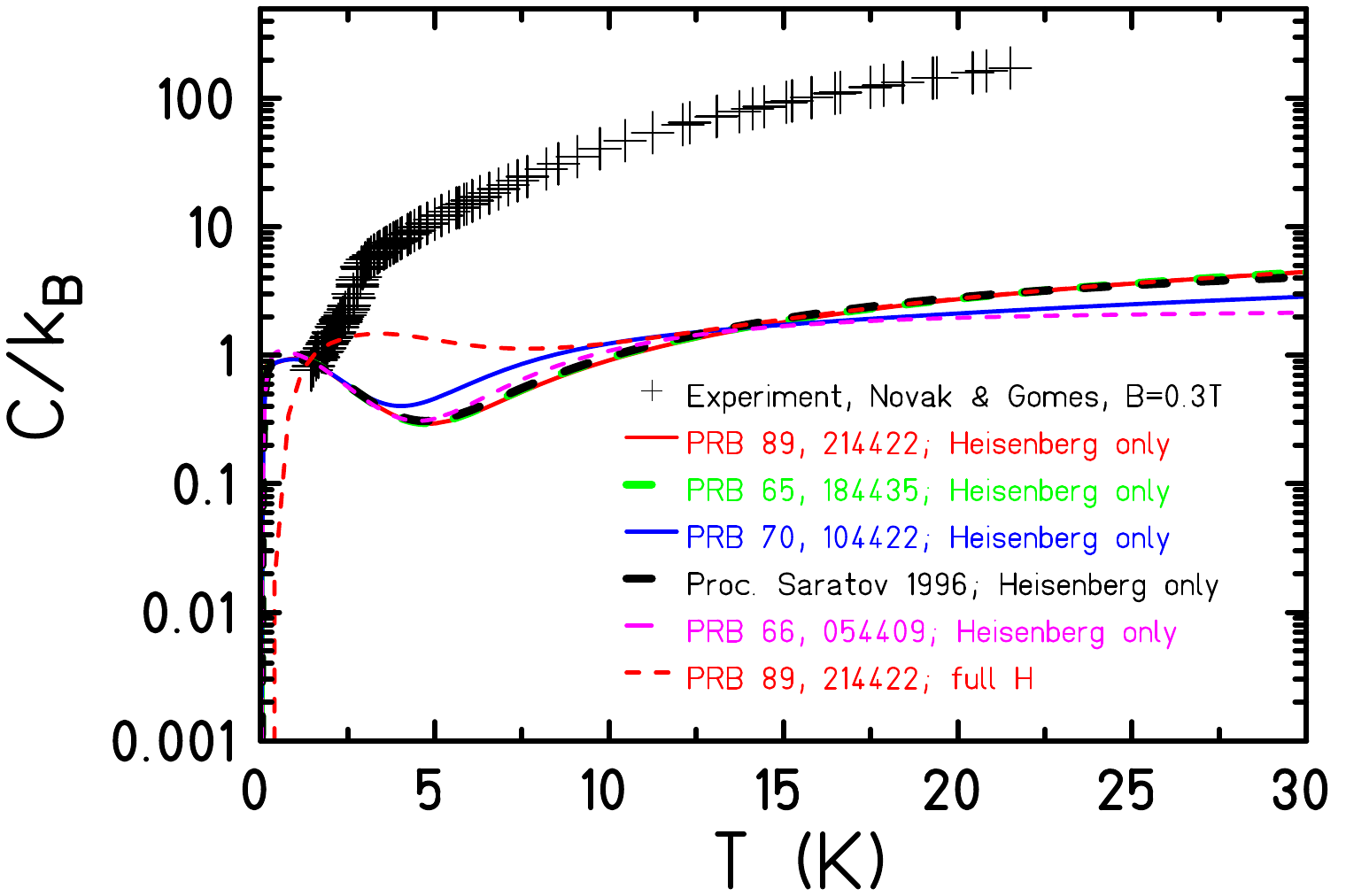}
\caption{(Color online) Specific heat of \MnAC\ at $B=0$ (top)
  and $B=0.3$~T 
  (bottom). Data taken from Ref.~\onlinecite{GNS:PRB98}. Color
  code of curves as above. For $B=0.3$~T the calculation for the
  anisotropic spin model was averaged over 20 directions.}   
\label{tlmm-f-g}
\end{figure}

Another observable that was measured very early in the history
of \MnAC\ is the heat
capacity.\cite{GNS:PRB98,FLB:PRL98,LMT:PRL00}
Figure~\xref{tlmm-f-g} shows the experimental data of
Ref.~\onlinecite{GNS:PRB98} for $B=0$ (top) and $B=0.3$~T
(bottom). One notices that the heat capacity is rather large and
grows steadily with temperature. This is due to a massive
contribution from lattice vibrations (phonons) which grows like
$T^3$. Therefore, heat capacity data of magnetic molecules are
usually overwhelmed by phonon contributions above $T=5$~K.

This fact becomes obvious when comparing the theoretical heat
capacity data in \figref{tlmm-f-g} for the Heisenberg
parameterizations at $B=0$. At low temperatures the theoretical
values are two or more orders of magnitude smaller than the
experimental ones. But for the calculation using the full
anisotropic Hamiltonian\cite{MKJ:PRB14} one notices that the low-temperature
values agree very nicely. We think that this is due to a more
smeared-out density of states at low energies in the anisotropic
model whereas for Heisenberg systems these levels belong to
highly degenerate multiplets which leads to a different,
i.e. much smaller heat capacity. 

Interestingly, a magnetic field of $B=0.3$~T has a similar
effect. It smears out the density of states due to Zeeman
splitting. Therefore even for the plain Heisenberg models the
low-temperature heat capacity increases, but still does not
agree with the experimental data. For the anisotropic
spin-Hamiltonian\cite{MKJ:PRB14} the low-temperature heat
capacity does not change much and still agrees nicely with the
experimental data. We conjecture that although the energy levels
are moved around by the magnetic field, the overall structure of
the density of states remains very similar.

Summarizing, the specific heat is well reproduced by the
anisotropic spin-Hamiltonian of Ref.~\onlinecite{MKJ:PRB14} for
low-temperatures around 1~K.

\section{Summary and Outlook}
\label{sec-6}

35 years after its synthesis and 22 years after the first
measurements\cite{STS:JACS93} of \MnAC\ the 
Finite-Temperature Lanczos Method puts us in a position to
evaluate thermodynamic functions of really large magnetic
molecules. It thus complements DFT calculations of such big
systems in so far that one does no longer need to stop half way
to an understanding of thermodynamic observables. The next major
necessary step is now to develop tools for an evaluation of
non-equilibrium properties of such big quantum spin systems.

\section*{Acknowledgment}

This work was supported by the German Science Foundation (DFG
SCHN 615/15-1). Computing time at the Leibniz
Computing Center in Garching is also gratefully
acknowledged. We thank Roberta Sessoli, Mark Murrie, Thorsten
Glaser, Stephan Walleck, Miguel Novak and Angelo Gomes for
providing their very valuable experimental data 
for comparison and we thank Alexander Lichtenstein,
Mikhail Katsnelson as well as Roberta Sessoli for fruitful
discussions.

%


\begin{thebibliography}{56}%
\makeatletter
\providecommand \@ifxundefined [1]{%
 \@ifx{#1\undefined}
}%
\providecommand \@ifnum [1]{%
 \ifnum #1\expandafter \@firstoftwo
 \else \expandafter \@secondoftwo
 \fi
}%
\providecommand \@ifx [1]{%
 \ifx #1\expandafter \@firstoftwo
 \else \expandafter \@secondoftwo
 \fi
}%
\providecommand \natexlab [1]{#1}%
\providecommand \enquote  [1]{``#1''}%
\providecommand \bibnamefont  [1]{#1}%
\providecommand \bibfnamefont [1]{#1}%
\providecommand \citenamefont [1]{#1}%
\providecommand \href@noop [0]{\@secondoftwo}%
\providecommand \href [0]{\begingroup \@sanitize@url \@href}%
\providecommand \@href[1]{\@@startlink{#1}\@@href}%
\providecommand \@@href[1]{\endgroup#1\@@endlink}%
\providecommand \@sanitize@url [0]{\catcode `\\12\catcode `\$12\catcode
  `\&12\catcode `\#12\catcode `\^12\catcode `\_12\catcode `\%12\relax}%
\providecommand \@@startlink[1]{}%
\providecommand \@@endlink[0]{}%
\providecommand \url  [0]{\begingroup\@sanitize@url \@url }%
\providecommand \@url [1]{\endgroup\@href {#1}{\urlprefix }}%
\providecommand \urlprefix  [0]{URL }%
\providecommand \Eprint [0]{\href }%
\providecommand \doibase [0]{http://dx.doi.org/}%
\providecommand \selectlanguage [0]{\@gobble}%
\providecommand \bibinfo  [0]{\@secondoftwo}%
\providecommand \bibfield  [0]{\@secondoftwo}%
\providecommand \translation [1]{[#1]}%
\providecommand \BibitemOpen [0]{}%
\providecommand \bibitemStop [0]{}%
\providecommand \bibitemNoStop [0]{.\EOS\space}%
\providecommand \EOS [0]{\spacefactor3000\relax}%
\providecommand \BibitemShut  [1]{\csname bibitem#1\endcsname}%
\let\auto@bib@innerbib\@empty
\bibitem [{\citenamefont {Liechtenstein}\ \emph {et~al.}(1987)\citenamefont
  {Liechtenstein}, \citenamefont {Katsnelson}, \citenamefont {Antropov},\ and\
  \citenamefont {Gubanov}}]{LKA:JMMM87}%
  \BibitemOpen
  \bibfield  {author} {\bibinfo {author} {\bibfnamefont {A.}~\bibnamefont
  {Liechtenstein}}, \bibinfo {author} {\bibfnamefont {M.}~\bibnamefont
  {Katsnelson}}, \bibinfo {author} {\bibfnamefont {V.}~\bibnamefont
  {Antropov}}, \ and\ \bibinfo {author} {\bibfnamefont {V.}~\bibnamefont
  {Gubanov}},\ }\href {\doibase http://dx.doi.org/10.1016/0304-8853(87)90721-9}
  {\bibfield  {journal} {\bibinfo  {journal} {J. Magn. Magn. Mater.}\ }\textbf
  {\bibinfo {volume} {67}},\ \bibinfo {pages} {65 } (\bibinfo {year}
  {1987})}\BibitemShut {NoStop}%
\bibitem [{\citenamefont {Ruiz}\ \emph {et~al.}(1997)\citenamefont {Ruiz},
  \citenamefont {Alemany}, \citenamefont {Alvarez},\ and\ \citenamefont
  {Cano}}]{RAA:JACS97}%
  \BibitemOpen
  \bibfield  {author} {\bibinfo {author} {\bibfnamefont {E.}~\bibnamefont
  {Ruiz}}, \bibinfo {author} {\bibfnamefont {P.}~\bibnamefont {Alemany}},
  \bibinfo {author} {\bibfnamefont {S.}~\bibnamefont {Alvarez}}, \ and\
  \bibinfo {author} {\bibfnamefont {J.}~\bibnamefont {Cano}},\ }\href@noop {}
  {\bibfield  {journal} {\bibinfo  {journal} {J. Am. Chem. Soc.}\ }\textbf
  {\bibinfo {volume} {119}},\ \bibinfo {pages} {1297} (\bibinfo {year}
  {1997})}\BibitemShut {NoStop}%
\bibitem [{\citenamefont {Kortus}\ \emph {et~al.}(2001)\citenamefont {Kortus},
  \citenamefont {Hellberg},\ and\ \citenamefont {Pederson}}]{KHP:PRL01}%
  \BibitemOpen
  \bibfield  {author} {\bibinfo {author} {\bibfnamefont {J.}~\bibnamefont
  {Kortus}}, \bibinfo {author} {\bibfnamefont {C.~S.}\ \bibnamefont
  {Hellberg}}, \ and\ \bibinfo {author} {\bibfnamefont {M.~R.}\ \bibnamefont
  {Pederson}},\ }\href@noop {} {\bibfield  {journal} {\bibinfo  {journal}
  {Phys. Rev. Lett.}\ }\textbf {\bibinfo {volume} {86}},\ \bibinfo {pages}
  {3400} (\bibinfo {year} {2001})}\BibitemShut {NoStop}%
\bibitem [{\citenamefont {De~Raedt}\ \emph {et~al.}(2002)\citenamefont
  {De~Raedt}, \citenamefont {Hams}, \citenamefont {Dobrovitski}, \citenamefont
  {Al-Saqer}, \citenamefont {Katsnelson},\ and\ \citenamefont
  {Harmon}}]{RHD:JMMM02}%
  \BibitemOpen
  \bibfield  {author} {\bibinfo {author} {\bibfnamefont {H.~A.}\ \bibnamefont
  {De~Raedt}}, \bibinfo {author} {\bibfnamefont {A.~H.}\ \bibnamefont {Hams}},
  \bibinfo {author} {\bibfnamefont {V.~V.}\ \bibnamefont {Dobrovitski}},
  \bibinfo {author} {\bibfnamefont {M.}~\bibnamefont {Al-Saqer}}, \bibinfo
  {author} {\bibfnamefont {M.~I.}\ \bibnamefont {Katsnelson}}, \ and\ \bibinfo
  {author} {\bibfnamefont {B.~N.}\ \bibnamefont {Harmon}},\ }\href@noop {}
  {\bibfield  {journal} {\bibinfo  {journal} {J. Magn. Magn. Mater.}\ }\textbf
  {\bibinfo {volume} {246}},\ \bibinfo {pages} {392} (\bibinfo {year}
  {2002})}\BibitemShut {NoStop}%
\bibitem [{\citenamefont {Boukhvalov}\ \emph {et~al.}(2003)\citenamefont
  {Boukhvalov}, \citenamefont {Kurmaev}, \citenamefont {Moewes}, \citenamefont
  {Zatsepin}, \citenamefont {Cherkashenko}, \citenamefont {Nemnonov},
  \citenamefont {Finkelstein}, \citenamefont {Yarmoshenko}, \citenamefont
  {Neumann}, \citenamefont {Dobrovitski}, \citenamefont {Katsnelson},
  \citenamefont {Lichtenstein}, \citenamefont {Harmon},\ and\ \citenamefont
  {K{\"o}gerler}}]{BKM:PRB03}%
  \BibitemOpen
  \bibfield  {author} {\bibinfo {author} {\bibfnamefont {D.~W.}\ \bibnamefont
  {Boukhvalov}}, \bibinfo {author} {\bibfnamefont {E.~Z.}\ \bibnamefont
  {Kurmaev}}, \bibinfo {author} {\bibfnamefont {A.}~\bibnamefont {Moewes}},
  \bibinfo {author} {\bibfnamefont {D.~A.}\ \bibnamefont {Zatsepin}}, \bibinfo
  {author} {\bibfnamefont {V.~M.}\ \bibnamefont {Cherkashenko}}, \bibinfo
  {author} {\bibfnamefont {S.~N.}\ \bibnamefont {Nemnonov}}, \bibinfo {author}
  {\bibfnamefont {L.~D.}\ \bibnamefont {Finkelstein}}, \bibinfo {author}
  {\bibfnamefont {Y.~M.}\ \bibnamefont {Yarmoshenko}}, \bibinfo {author}
  {\bibfnamefont {M.}~\bibnamefont {Neumann}}, \bibinfo {author} {\bibfnamefont
  {V.~V.}\ \bibnamefont {Dobrovitski}}, \bibinfo {author} {\bibfnamefont
  {M.~I.}\ \bibnamefont {Katsnelson}}, \bibinfo {author} {\bibfnamefont
  {A.~I.}\ \bibnamefont {Lichtenstein}}, \bibinfo {author} {\bibfnamefont
  {B.~N.}\ \bibnamefont {Harmon}}, \ and\ \bibinfo {author} {\bibfnamefont
  {P.}~\bibnamefont {K{\"o}gerler}},\ }\href@noop {} {\bibfield  {journal}
  {\bibinfo  {journal} {Phys. Rev. B}\ }\textbf {\bibinfo {volume} {67}},\
  \bibinfo {pages} {134408} (\bibinfo {year} {2003})}\BibitemShut {NoStop}%
\bibitem [{\citenamefont {Baruah}\ \emph {et~al.}(2004)\citenamefont {Baruah},
  \citenamefont {Kortus}, \citenamefont {Pederson}, \citenamefont {Wesolowski},
  \citenamefont {Haraldsen}, \citenamefont {Musfeldt}, \citenamefont {North},
  \citenamefont {Zipse},\ and\ \citenamefont {Dalal}}]{BKP:PRB2004}%
  \BibitemOpen
  \bibfield  {author} {\bibinfo {author} {\bibfnamefont {T.}~\bibnamefont
  {Baruah}}, \bibinfo {author} {\bibfnamefont {J.}~\bibnamefont {Kortus}},
  \bibinfo {author} {\bibfnamefont {M.~R.}\ \bibnamefont {Pederson}}, \bibinfo
  {author} {\bibfnamefont {R.}~\bibnamefont {Wesolowski}}, \bibinfo {author}
  {\bibfnamefont {J.~T.}\ \bibnamefont {Haraldsen}}, \bibinfo {author}
  {\bibfnamefont {J.~L.}\ \bibnamefont {Musfeldt}}, \bibinfo {author}
  {\bibfnamefont {J.~M.}\ \bibnamefont {North}}, \bibinfo {author}
  {\bibfnamefont {D.}~\bibnamefont {Zipse}}, \ and\ \bibinfo {author}
  {\bibfnamefont {N.~S.}\ \bibnamefont {Dalal}},\ }\href@noop {} {\bibfield
  {journal} {\bibinfo  {journal} {Phys. Rev. B}\ }\textbf {\bibinfo {volume}
  {70}},\ \bibinfo {pages} {214410} (\bibinfo {year} {2004})}\BibitemShut
  {NoStop}%
\bibitem [{\citenamefont {Zaharko}\ \emph {et~al.}(2008)\citenamefont
  {Zaharko}, \citenamefont {Mesot}, \citenamefont {Salguero}, \citenamefont
  {Valent\'{\i}}, \citenamefont {Zbiri}, \citenamefont {Johnson}, \citenamefont
  {Filinchuk}, \citenamefont {Klemke}, \citenamefont {Kiefer}, \citenamefont
  {Mys'kiv}, \citenamefont {Str\"{a}ssle},\ and\ \citenamefont
  {Mutka}}]{ZMS:PRB08A}%
  \BibitemOpen
  \bibfield  {author} {\bibinfo {author} {\bibfnamefont {O.}~\bibnamefont
  {Zaharko}}, \bibinfo {author} {\bibfnamefont {J.}~\bibnamefont {Mesot}},
  \bibinfo {author} {\bibfnamefont {L.~A.}\ \bibnamefont {Salguero}}, \bibinfo
  {author} {\bibfnamefont {R.}~\bibnamefont {Valent\'{\i}}}, \bibinfo {author}
  {\bibfnamefont {M.}~\bibnamefont {Zbiri}}, \bibinfo {author} {\bibfnamefont
  {M.}~\bibnamefont {Johnson}}, \bibinfo {author} {\bibfnamefont
  {Y.}~\bibnamefont {Filinchuk}}, \bibinfo {author} {\bibfnamefont
  {B.}~\bibnamefont {Klemke}}, \bibinfo {author} {\bibfnamefont
  {K.}~\bibnamefont {Kiefer}}, \bibinfo {author} {\bibfnamefont
  {M.}~\bibnamefont {Mys'kiv}}, \bibinfo {author} {\bibfnamefont
  {T.}~\bibnamefont {Str\"{a}ssle}}, \ and\ \bibinfo {author} {\bibfnamefont
  {H.}~\bibnamefont {Mutka}},\ }\href {\doibase 10.1103/PhysRevB.77.224408}
  {\bibfield  {journal} {\bibinfo  {journal} {Phys. Rev. B}\ }\textbf {\bibinfo
  {volume} {77}},\ \bibinfo {eid} {224408} (\bibinfo {year}
  {2008})}\BibitemShut {NoStop}%
\bibitem [{\citenamefont {Zartilas}\ \emph {et~al.}(2013)\citenamefont
  {Zartilas}, \citenamefont {Papatriantafyllopoulou}, \citenamefont
  {Stamatatos}, \citenamefont {Nastopoulos}, \citenamefont {Cremades},
  \citenamefont {Ruiz}, \citenamefont {Christou}, \citenamefont
  {Lampropoulos},\ and\ \citenamefont {Tasiopoulos}}]{ZPS:IC13}%
  \BibitemOpen
  \bibfield  {author} {\bibinfo {author} {\bibfnamefont {S.}~\bibnamefont
  {Zartilas}}, \bibinfo {author} {\bibfnamefont {C.}~\bibnamefont
  {Papatriantafyllopoulou}}, \bibinfo {author} {\bibfnamefont {T.~C.}\
  \bibnamefont {Stamatatos}}, \bibinfo {author} {\bibfnamefont
  {V.}~\bibnamefont {Nastopoulos}}, \bibinfo {author} {\bibfnamefont
  {E.}~\bibnamefont {Cremades}}, \bibinfo {author} {\bibfnamefont
  {E.}~\bibnamefont {Ruiz}}, \bibinfo {author} {\bibfnamefont {G.}~\bibnamefont
  {Christou}}, \bibinfo {author} {\bibfnamefont {C.}~\bibnamefont
  {Lampropoulos}}, \ and\ \bibinfo {author} {\bibfnamefont {A.~J.}\
  \bibnamefont {Tasiopoulos}},\ }\href {\doibase 10.1021/ic401872c} {\bibfield
  {journal} {\bibinfo  {journal} {Inorg. Chem.}\ }\textbf {\bibinfo {volume}
  {52}},\ \bibinfo {pages} {12070} (\bibinfo {year} {2013})}\BibitemShut
  {NoStop}%
\bibitem [{\citenamefont {Kuepper}\ \emph {et~al.}(2013)\citenamefont
  {Kuepper}, \citenamefont {Derks}, \citenamefont {Taubitz}, \citenamefont
  {Prinz}, \citenamefont {Joly}, \citenamefont {Kappler}, \citenamefont
  {Postnikov}, \citenamefont {Yang}, \citenamefont {Kuznetsova}, \citenamefont
  {Wiedwald}, \citenamefont {Ziemann},\ and\ \citenamefont
  {Neumann}}]{KDT:DT13}%
  \BibitemOpen
  \bibfield  {author} {\bibinfo {author} {\bibfnamefont {K.}~\bibnamefont
  {Kuepper}}, \bibinfo {author} {\bibfnamefont {C.}~\bibnamefont {Derks}},
  \bibinfo {author} {\bibfnamefont {C.}~\bibnamefont {Taubitz}}, \bibinfo
  {author} {\bibfnamefont {M.}~\bibnamefont {Prinz}}, \bibinfo {author}
  {\bibfnamefont {L.}~\bibnamefont {Joly}}, \bibinfo {author} {\bibfnamefont
  {J.-P.}\ \bibnamefont {Kappler}}, \bibinfo {author} {\bibfnamefont
  {A.}~\bibnamefont {Postnikov}}, \bibinfo {author} {\bibfnamefont
  {W.}~\bibnamefont {Yang}}, \bibinfo {author} {\bibfnamefont {T.~V.}\
  \bibnamefont {Kuznetsova}}, \bibinfo {author} {\bibfnamefont
  {U.}~\bibnamefont {Wiedwald}}, \bibinfo {author} {\bibfnamefont
  {P.}~\bibnamefont {Ziemann}}, \ and\ \bibinfo {author} {\bibfnamefont
  {M.}~\bibnamefont {Neumann}},\ }\href {\doibase 10.1039/C3DT32759K}
  {\bibfield  {journal} {\bibinfo  {journal} {Dalton Trans.}\ }\textbf
  {\bibinfo {volume} {42}},\ \bibinfo {pages} {7924} (\bibinfo {year}
  {2013})}\BibitemShut {NoStop}%
\bibitem [{\citenamefont {Chiesa}\ \emph {et~al.}(2013)\citenamefont {Chiesa},
  \citenamefont {Carretta}, \citenamefont {Santini}, \citenamefont {Amoretti},\
  and\ \citenamefont {Pavarini}}]{CCS:PRL13}%
  \BibitemOpen
  \bibfield  {author} {\bibinfo {author} {\bibfnamefont {A.}~\bibnamefont
  {Chiesa}}, \bibinfo {author} {\bibfnamefont {S.}~\bibnamefont {Carretta}},
  \bibinfo {author} {\bibfnamefont {P.}~\bibnamefont {Santini}}, \bibinfo
  {author} {\bibfnamefont {G.}~\bibnamefont {Amoretti}}, \ and\ \bibinfo
  {author} {\bibfnamefont {E.}~\bibnamefont {Pavarini}},\ }\href {\doibase
  10.1103/PhysRevLett.110.157204} {\bibfield  {journal} {\bibinfo  {journal}
  {Phys. Rev. Lett.}\ }\textbf {\bibinfo {volume} {110}},\ \bibinfo {pages}
  {157204} (\bibinfo {year} {2013})}\BibitemShut {NoStop}%
\bibitem [{\citenamefont {Silverstein}\ \emph {et~al.}(2014)\citenamefont
  {Silverstein}, \citenamefont {Fritsch}, \citenamefont {Flicker},
  \citenamefont {Hallas}, \citenamefont {Gardner}, \citenamefont {Qiu},
  \citenamefont {Ehlers}, \citenamefont {Savici}, \citenamefont {Yamani},
  \citenamefont {Ross}, \citenamefont {Gaulin}, \citenamefont {Gingras},
  \citenamefont {Paddison}, \citenamefont {Foyevtsova}, \citenamefont
  {Valenti}, \citenamefont {Hawthorne}, \citenamefont {Wiebe},\ and\
  \citenamefont {Zhou}}]{SFF:PRB14}%
  \BibitemOpen
  \bibfield  {author} {\bibinfo {author} {\bibfnamefont {H.~J.}\ \bibnamefont
  {Silverstein}}, \bibinfo {author} {\bibfnamefont {K.}~\bibnamefont
  {Fritsch}}, \bibinfo {author} {\bibfnamefont {F.}~\bibnamefont {Flicker}},
  \bibinfo {author} {\bibfnamefont {A.~M.}\ \bibnamefont {Hallas}}, \bibinfo
  {author} {\bibfnamefont {J.~S.}\ \bibnamefont {Gardner}}, \bibinfo {author}
  {\bibfnamefont {Y.}~\bibnamefont {Qiu}}, \bibinfo {author} {\bibfnamefont
  {G.}~\bibnamefont {Ehlers}}, \bibinfo {author} {\bibfnamefont {A.~T.}\
  \bibnamefont {Savici}}, \bibinfo {author} {\bibfnamefont {Z.}~\bibnamefont
  {Yamani}}, \bibinfo {author} {\bibfnamefont {K.~A.}\ \bibnamefont {Ross}},
  \bibinfo {author} {\bibfnamefont {B.~D.}\ \bibnamefont {Gaulin}}, \bibinfo
  {author} {\bibfnamefont {M.~J.~P.}\ \bibnamefont {Gingras}}, \bibinfo
  {author} {\bibfnamefont {J.~A.~M.}\ \bibnamefont {Paddison}}, \bibinfo
  {author} {\bibfnamefont {K.}~\bibnamefont {Foyevtsova}}, \bibinfo {author}
  {\bibfnamefont {R.}~\bibnamefont {Valenti}}, \bibinfo {author} {\bibfnamefont
  {F.}~\bibnamefont {Hawthorne}}, \bibinfo {author} {\bibfnamefont {C.~R.}\
  \bibnamefont {Wiebe}}, \ and\ \bibinfo {author} {\bibfnamefont {H.~D.}\
  \bibnamefont {Zhou}},\ }\href {\doibase 10.1103/PhysRevB.89.054433}
  {\bibfield  {journal} {\bibinfo  {journal} {Phys. Rev. B}\ }\textbf {\bibinfo
  {volume} {89}},\ \bibinfo {pages} {054433} (\bibinfo {year}
  {2014})}\BibitemShut {NoStop}%
\bibitem [{\citenamefont {Pedersen}\ \emph {et~al.}(2014)\citenamefont
  {Pedersen}, \citenamefont {Lorusso}, \citenamefont {Morales}, \citenamefont
  {Weyherm{\"u}ller}, \citenamefont {Piligkos}, \citenamefont {Singh},
  \citenamefont {Larsen}, \citenamefont {Schau-Magnussen}, \citenamefont
  {Rajaraman}, \citenamefont {Evangelisti},\ and\ \citenamefont
  {Bendix}}]{PLM:ACIE14}%
  \BibitemOpen
  \bibfield  {author} {\bibinfo {author} {\bibfnamefont {K.~S.}\ \bibnamefont
  {Pedersen}}, \bibinfo {author} {\bibfnamefont {G.}~\bibnamefont {Lorusso}},
  \bibinfo {author} {\bibfnamefont {J.~J.}\ \bibnamefont {Morales}}, \bibinfo
  {author} {\bibfnamefont {T.}~\bibnamefont {Weyherm{\"u}ller}}, \bibinfo
  {author} {\bibfnamefont {S.}~\bibnamefont {Piligkos}}, \bibinfo {author}
  {\bibfnamefont {S.~K.}\ \bibnamefont {Singh}}, \bibinfo {author}
  {\bibfnamefont {D.}~\bibnamefont {Larsen}}, \bibinfo {author} {\bibfnamefont
  {M.}~\bibnamefont {Schau-Magnussen}}, \bibinfo {author} {\bibfnamefont
  {G.}~\bibnamefont {Rajaraman}}, \bibinfo {author} {\bibfnamefont
  {M.}~\bibnamefont {Evangelisti}}, \ and\ \bibinfo {author} {\bibfnamefont
  {J.}~\bibnamefont {Bendix}},\ }\href {\doibase 10.1002/anie.201308240}
  {\bibfield  {journal} {\bibinfo  {journal} {Angew. Chem. Int. Ed.}\ }\textbf
  {\bibinfo {volume} {53}},\ \bibinfo {pages} {2394} (\bibinfo {year}
  {2014})}\BibitemShut {NoStop}%
\bibitem [{\citenamefont {Sanz}\ \emph {et~al.}(2014)\citenamefont {Sanz},
  \citenamefont {Frost}, \citenamefont {Rajeshkumar}, \citenamefont {Dalgarno},
  \citenamefont {Rajaraman}, \citenamefont {Wernsdorfer}, \citenamefont
  {Schnack}, \citenamefont {Lusby},\ and\ \citenamefont
  {Brechin}}]{SFR:CAEJ14}%
  \BibitemOpen
  \bibfield  {author} {\bibinfo {author} {\bibfnamefont {S.}~\bibnamefont
  {Sanz}}, \bibinfo {author} {\bibfnamefont {J.~M.}\ \bibnamefont {Frost}},
  \bibinfo {author} {\bibfnamefont {T.}~\bibnamefont {Rajeshkumar}}, \bibinfo
  {author} {\bibfnamefont {S.~J.}\ \bibnamefont {Dalgarno}}, \bibinfo {author}
  {\bibfnamefont {G.}~\bibnamefont {Rajaraman}}, \bibinfo {author}
  {\bibfnamefont {W.}~\bibnamefont {Wernsdorfer}}, \bibinfo {author}
  {\bibfnamefont {J.}~\bibnamefont {Schnack}}, \bibinfo {author} {\bibfnamefont
  {P.~J.}\ \bibnamefont {Lusby}}, \ and\ \bibinfo {author} {\bibfnamefont
  {E.~K.}\ \bibnamefont {Brechin}},\ }\href
  {http://dx.doi.org/10.1002/chem.201304740} {\bibfield  {journal} {\bibinfo
  {journal} {Chem. Eur. J.}\ }\textbf {\bibinfo {volume} {20}},\ \bibinfo
  {pages} {3010} (\bibinfo {year} {2014})}\BibitemShut {NoStop}%
\bibitem [{\citenamefont {Singh}\ and\ \citenamefont
  {Rajaraman}(2014)}]{SiR:CAEJ14}%
  \BibitemOpen
  \bibfield  {author} {\bibinfo {author} {\bibfnamefont {S.~K.}\ \bibnamefont
  {Singh}}\ and\ \bibinfo {author} {\bibfnamefont {G.}~\bibnamefont
  {Rajaraman}},\ }\href {\doibase 10.1002/chem.201303489} {\bibfield  {journal}
  {\bibinfo  {journal} {Chem. Eur. J.}\ }\textbf {\bibinfo {volume} {20}},\
  \bibinfo {pages} {113} (\bibinfo {year} {2014})}\BibitemShut {NoStop}%
\bibitem [{\citenamefont {Kvashnin}\ \emph {et~al.}(2015)\citenamefont
  {Kvashnin}, \citenamefont {Gr\aa{}n\"as}, \citenamefont {Di~Marco},
  \citenamefont {Katsnelson}, \citenamefont {Lichtenstein},\ and\ \citenamefont
  {Eriksson}}]{KGD:PRB15}%
  \BibitemOpen
  \bibfield  {author} {\bibinfo {author} {\bibfnamefont {Y.~O.}\ \bibnamefont
  {Kvashnin}}, \bibinfo {author} {\bibfnamefont {O.}~\bibnamefont
  {Gr\aa{}n\"as}}, \bibinfo {author} {\bibfnamefont {I.}~\bibnamefont
  {Di~Marco}}, \bibinfo {author} {\bibfnamefont {M.~I.}\ \bibnamefont
  {Katsnelson}}, \bibinfo {author} {\bibfnamefont {A.~I.}\ \bibnamefont
  {Lichtenstein}}, \ and\ \bibinfo {author} {\bibfnamefont {O.}~\bibnamefont
  {Eriksson}},\ }\href {\doibase 10.1103/PhysRevB.91.125133} {\bibfield
  {journal} {\bibinfo  {journal} {Phys. Rev. B}\ }\textbf {\bibinfo {volume}
  {91}},\ \bibinfo {pages} {125133} (\bibinfo {year} {2015})}\BibitemShut
  {NoStop}%
\bibitem [{\citenamefont {Mazurenko}\ \emph {et~al.}(2014)\citenamefont
  {Mazurenko}, \citenamefont {Kvashnin}, \citenamefont {Jin}, \citenamefont
  {De~Raedt}, \citenamefont {Lichtenstein},\ and\ \citenamefont
  {Katsnelson}}]{MKJ:PRB14}%
  \BibitemOpen
  \bibfield  {author} {\bibinfo {author} {\bibfnamefont {V.~V.}\ \bibnamefont
  {Mazurenko}}, \bibinfo {author} {\bibfnamefont {Y.~O.}\ \bibnamefont
  {Kvashnin}}, \bibinfo {author} {\bibfnamefont {F.}~\bibnamefont {Jin}},
  \bibinfo {author} {\bibfnamefont {H.~A.}\ \bibnamefont {De~Raedt}}, \bibinfo
  {author} {\bibfnamefont {A.~I.}\ \bibnamefont {Lichtenstein}}, \ and\
  \bibinfo {author} {\bibfnamefont {M.~I.}\ \bibnamefont {Katsnelson}},\ }\href
  {\doibase 10.1103/PhysRevB.89.214422} {\bibfield  {journal} {\bibinfo
  {journal} {Phys. Rev. B}\ }\textbf {\bibinfo {volume} {89}},\ \bibinfo
  {pages} {214422} (\bibinfo {year} {2014})}\BibitemShut {NoStop}%
\bibitem [{\citenamefont {Kortus}\ \emph {et~al.}(2002)\citenamefont {Kortus},
  \citenamefont {Baruah}, \citenamefont {Bernstein},\ and\ \citenamefont
  {Pederson}}]{KBB:PRB02}%
  \BibitemOpen
  \bibfield  {author} {\bibinfo {author} {\bibfnamefont {J.}~\bibnamefont
  {Kortus}}, \bibinfo {author} {\bibfnamefont {T.}~\bibnamefont {Baruah}},
  \bibinfo {author} {\bibfnamefont {N.}~\bibnamefont {Bernstein}}, \ and\
  \bibinfo {author} {\bibfnamefont {M.~R.}\ \bibnamefont {Pederson}},\
  }\href@noop {} {\bibfield  {journal} {\bibinfo  {journal} {Phys. Rev. B}\
  }\textbf {\bibinfo {volume} {66}},\ \bibinfo {pages} {092403} (\bibinfo
  {year} {2002})}\BibitemShut {NoStop}%
\bibitem [{\citenamefont {Boukhvalov}\ \emph {et~al.}(2002)\citenamefont
  {Boukhvalov}, \citenamefont {Lichtenstein}, \citenamefont {Dobrovitski},
  \citenamefont {Katsnelson}, \citenamefont {Harmon}, \citenamefont
  {Mazurenko},\ and\ \citenamefont {Anisimov}}]{BLD:PRB02}%
  \BibitemOpen
  \bibfield  {author} {\bibinfo {author} {\bibfnamefont {D.~W.}\ \bibnamefont
  {Boukhvalov}}, \bibinfo {author} {\bibfnamefont {A.~I.}\ \bibnamefont
  {Lichtenstein}}, \bibinfo {author} {\bibfnamefont {V.~V.}\ \bibnamefont
  {Dobrovitski}}, \bibinfo {author} {\bibfnamefont {M.~I.}\ \bibnamefont
  {Katsnelson}}, \bibinfo {author} {\bibfnamefont {B.~N.}\ \bibnamefont
  {Harmon}}, \bibinfo {author} {\bibfnamefont {V.~V.}\ \bibnamefont
  {Mazurenko}}, \ and\ \bibinfo {author} {\bibfnamefont {V.~I.}\ \bibnamefont
  {Anisimov}},\ }\href {\doibase 10.1103/PhysRevB.65.184435} {\bibfield
  {journal} {\bibinfo  {journal} {Phys. Rev. B}\ }\textbf {\bibinfo {volume}
  {65}},\ \bibinfo {pages} {184435} (\bibinfo {year} {2002})}\BibitemShut
  {NoStop}%
\bibitem [{\citenamefont {Chaboussant}\ \emph {et~al.}(2004)\citenamefont
  {Chaboussant}, \citenamefont {Sieber}, \citenamefont {Ochsenbein},
  \citenamefont {G\"udel}, \citenamefont {Murrie}, \citenamefont {Honecker},
  \citenamefont {Fukushima},\ and\ \citenamefont {Normand}}]{CSO:PRB04}%
  \BibitemOpen
  \bibfield  {author} {\bibinfo {author} {\bibfnamefont {G.}~\bibnamefont
  {Chaboussant}}, \bibinfo {author} {\bibfnamefont {A.}~\bibnamefont {Sieber}},
  \bibinfo {author} {\bibfnamefont {S.}~\bibnamefont {Ochsenbein}}, \bibinfo
  {author} {\bibfnamefont {H.-U.}\ \bibnamefont {G\"udel}}, \bibinfo {author}
  {\bibfnamefont {M.}~\bibnamefont {Murrie}}, \bibinfo {author} {\bibfnamefont
  {A.}~\bibnamefont {Honecker}}, \bibinfo {author} {\bibfnamefont
  {N.}~\bibnamefont {Fukushima}}, \ and\ \bibinfo {author} {\bibfnamefont
  {B.}~\bibnamefont {Normand}},\ }\href {\doibase 10.1103/PhysRevB.70.104422}
  {\bibfield  {journal} {\bibinfo  {journal} {Phys. Rev. B}\ }\textbf {\bibinfo
  {volume} {70}},\ \bibinfo {pages} {104422} (\bibinfo {year}
  {2004})}\BibitemShut {NoStop}%
\bibitem [{\citenamefont {Gatteschi}\ \emph {et~al.}(2006)\citenamefont
  {Gatteschi}, \citenamefont {Sessoli},\ and\ \citenamefont
  {Villain}}]{GSV:2006}%
  \BibitemOpen
  \bibfield  {author} {\bibinfo {author} {\bibfnamefont {D.}~\bibnamefont
  {Gatteschi}}, \bibinfo {author} {\bibfnamefont {R.}~\bibnamefont {Sessoli}},
  \ and\ \bibinfo {author} {\bibfnamefont {J.}~\bibnamefont {Villain}},\
  }\href@noop {} {\emph {\bibinfo {title} {Molecular Nanomagnets}}},\
  Mesoscopic Physics and Nanotechnology\ (\bibinfo  {publisher} {Oxford
  University Press},\ \bibinfo {address} {Oxford},\ \bibinfo {year}
  {2006})\BibitemShut {NoStop}%
\bibitem [{\citenamefont {Lis}(1980)}]{Lis:ACB80}%
  \BibitemOpen
  \bibfield  {author} {\bibinfo {author} {\bibfnamefont {T.}~\bibnamefont
  {Lis}},\ }\href {http://dx.doi.org/10.1140/epjb/e2010-00028-3} {\bibfield
  {journal} {\bibinfo  {journal} {Acta Chrytallogr. B}\ }\textbf {\bibinfo
  {volume} {36}},\ \bibinfo {pages} {2042} (\bibinfo {year}
  {1980})}\BibitemShut {NoStop}%
\bibitem [{\citenamefont {Sessoli}\ \emph
  {et~al.}(1993{\natexlab{a}})\citenamefont {Sessoli}, \citenamefont {Tsai},
  \citenamefont {Schake}, \citenamefont {Wang}, \citenamefont {Vincent},
  \citenamefont {Folting}, \citenamefont {Gatteschi}, \citenamefont
  {Christou},\ and\ \citenamefont {Hendrickson}}]{STS:JACS93}%
  \BibitemOpen
  \bibfield  {author} {\bibinfo {author} {\bibfnamefont {R.}~\bibnamefont
  {Sessoli}}, \bibinfo {author} {\bibfnamefont {H.~L.}\ \bibnamefont {Tsai}},
  \bibinfo {author} {\bibfnamefont {A.~R.}\ \bibnamefont {Schake}}, \bibinfo
  {author} {\bibfnamefont {S.}~\bibnamefont {Wang}}, \bibinfo {author}
  {\bibfnamefont {J.~B.}\ \bibnamefont {Vincent}}, \bibinfo {author}
  {\bibfnamefont {K.}~\bibnamefont {Folting}}, \bibinfo {author} {\bibfnamefont
  {D.}~\bibnamefont {Gatteschi}}, \bibinfo {author} {\bibfnamefont
  {G.}~\bibnamefont {Christou}}, \ and\ \bibinfo {author} {\bibfnamefont
  {D.~N.}\ \bibnamefont {Hendrickson}},\ }\href
  {http://pubs.acs.org/doi/abs/10.1021/ja00058a027} {\bibfield  {journal}
  {\bibinfo  {journal} {J. Am. Chem. Soc.}\ }\textbf {\bibinfo {volume}
  {115}},\ \bibinfo {pages} {1804} (\bibinfo {year}
  {1993}{\natexlab{a}})}\BibitemShut {NoStop}%
\bibitem [{\citenamefont {Sessoli}\ \emph
  {et~al.}(1993{\natexlab{b}})\citenamefont {Sessoli}, \citenamefont
  {Gatteschi}, \citenamefont {Caneschi},\ and\ \citenamefont
  {Novak}}]{SGC:Nat93}%
  \BibitemOpen
  \bibfield  {author} {\bibinfo {author} {\bibfnamefont {R.}~\bibnamefont
  {Sessoli}}, \bibinfo {author} {\bibfnamefont {D.}~\bibnamefont {Gatteschi}},
  \bibinfo {author} {\bibfnamefont {A.}~\bibnamefont {Caneschi}}, \ and\
  \bibinfo {author} {\bibfnamefont {M.~A.}\ \bibnamefont {Novak}},\ }\href
  {http://dx.doi.org/10.1038/365141a0} {\bibfield  {journal} {\bibinfo
  {journal} {Nature}\ }\textbf {\bibinfo {volume} {365}},\ \bibinfo {pages}
  {141} (\bibinfo {year} {1993}{\natexlab{b}})}\BibitemShut {NoStop}%
\bibitem [{\citenamefont {Thomas}\ \emph {et~al.}(1996)\citenamefont {Thomas},
  \citenamefont {Lionti}, \citenamefont {Ballou}, \citenamefont {Gatteschi},
  \citenamefont {Sessoli},\ and\ \citenamefont {Barbara}}]{TLB:Nature96}%
  \BibitemOpen
  \bibfield  {author} {\bibinfo {author} {\bibfnamefont {L.}~\bibnamefont
  {Thomas}}, \bibinfo {author} {\bibfnamefont {F.}~\bibnamefont {Lionti}},
  \bibinfo {author} {\bibfnamefont {R.}~\bibnamefont {Ballou}}, \bibinfo
  {author} {\bibfnamefont {D.}~\bibnamefont {Gatteschi}}, \bibinfo {author}
  {\bibfnamefont {R.}~\bibnamefont {Sessoli}}, \ and\ \bibinfo {author}
  {\bibfnamefont {B.}~\bibnamefont {Barbara}},\ }\href
  {http://dx.doi.org/10.1038/383145a0} {\bibfield  {journal} {\bibinfo
  {journal} {Nature}\ }\textbf {\bibinfo {volume} {383}},\ \bibinfo {pages}
  {145} (\bibinfo {year} {1996})}\BibitemShut {NoStop}%
\bibitem [{\citenamefont {Gomes}\ \emph {et~al.}(1998)\citenamefont {Gomes},
  \citenamefont {Novak}, \citenamefont {Sessoli}, \citenamefont {Caneschi},\
  and\ \citenamefont {Gatteschi}}]{GNS:PRB98}%
  \BibitemOpen
  \bibfield  {author} {\bibinfo {author} {\bibfnamefont {A.}~\bibnamefont
  {Gomes}}, \bibinfo {author} {\bibfnamefont {M.}~\bibnamefont {Novak}},
  \bibinfo {author} {\bibfnamefont {R.}~\bibnamefont {Sessoli}}, \bibinfo
  {author} {\bibfnamefont {A.}~\bibnamefont {Caneschi}}, \ and\ \bibinfo
  {author} {\bibfnamefont {D.}~\bibnamefont {Gatteschi}},\ }\href {\doibase
  10.1103/PhysRevB.57.5021} {\bibfield  {journal} {\bibinfo  {journal} {Phys.
  Rev. B}\ }\textbf {\bibinfo {volume} {57}},\ \bibinfo {pages} {5021}
  (\bibinfo {year} {1998})}\BibitemShut {NoStop}%
\bibitem [{\citenamefont {Cornia}\ \emph {et~al.}(2001)\citenamefont {Cornia},
  \citenamefont {Affronte}, \citenamefont {Gatteschi}, \citenamefont {Jansen},
  \citenamefont {Caneschi},\ and\ \citenamefont {Sessoli}}]{CAG:JMMM01}%
  \BibitemOpen
  \bibfield  {author} {\bibinfo {author} {\bibfnamefont {A.}~\bibnamefont
  {Cornia}}, \bibinfo {author} {\bibfnamefont {M.}~\bibnamefont {Affronte}},
  \bibinfo {author} {\bibfnamefont {A.~C. D.~T.}\ \bibnamefont {Gatteschi}},
  \bibinfo {author} {\bibfnamefont {A.~G.~M.}\ \bibnamefont {Jansen}}, \bibinfo
  {author} {\bibfnamefont {A.}~\bibnamefont {Caneschi}}, \ and\ \bibinfo
  {author} {\bibfnamefont {R.}~\bibnamefont {Sessoli}},\ }\href
  {http://dx.doi.org/10.1016/S0304-8853(00)01093-3} {\bibfield  {journal}
  {\bibinfo  {journal} {J. Magn. Magn. Mater.}\ }\textbf {\bibinfo {volume}
  {226}},\ \bibinfo {pages} {2012} (\bibinfo {year} {2001})}\BibitemShut
  {NoStop}%
\bibitem [{\citenamefont {Gatteschi}\ and\ \citenamefont
  {Sessoli}(2003)}]{GaS:ACIE03}%
  \BibitemOpen
  \bibfield  {author} {\bibinfo {author} {\bibfnamefont {D.}~\bibnamefont
  {Gatteschi}}\ and\ \bibinfo {author} {\bibfnamefont {R.}~\bibnamefont
  {Sessoli}},\ }\href {http://dx.doi.org/10.1002/anie.200390099} {\bibfield
  {journal} {\bibinfo  {journal} {Angew. Chem., Int. Edit.}\ }\textbf {\bibinfo
  {volume} {42}},\ \bibinfo {pages} {268} (\bibinfo {year} {2003})}\BibitemShut
  {NoStop}%
\bibitem [{\citenamefont {Schnalle}\ and\ \citenamefont
  {Schnack}(2009)}]{ScS:PRB09}%
  \BibitemOpen
  \bibfield  {author} {\bibinfo {author} {\bibfnamefont {R.}~\bibnamefont
  {Schnalle}}\ and\ \bibinfo {author} {\bibfnamefont {J.}~\bibnamefont
  {Schnack}},\ }\href {http://link.aps.org/abstract/PRB/v79/e104419} {\bibfield
   {journal} {\bibinfo  {journal} {Phys. Rev. B}\ }\textbf {\bibinfo {volume}
  {79}},\ \bibinfo {pages} {104419} (\bibinfo {year} {2009})}\BibitemShut
  {NoStop}%
\bibitem [{\citenamefont {Jakli\ifmmode~\check{c}\else \v{c}\fi{}}\ and\
  \citenamefont {Prelov\ifmmode~\check{s}\else
  \v{s}\fi{}ek}(1994)}]{PhysRevB.49.5065}%
  \BibitemOpen
  \bibfield  {author} {\bibinfo {author} {\bibfnamefont {J.}~\bibnamefont
  {Jakli\ifmmode~\check{c}\else \v{c}\fi{}}}\ and\ \bibinfo {author}
  {\bibfnamefont {P.}~\bibnamefont {Prelov\ifmmode~\check{s}\else
  \v{s}\fi{}ek}},\ }\href {http://dx.doi.org/10.1103/PhysRevB.49.5065}
  {\bibfield  {journal} {\bibinfo  {journal} {Phys. Rev. B}\ }\textbf {\bibinfo
  {volume} {49}},\ \bibinfo {pages} {5065} (\bibinfo {year}
  {1994})}\BibitemShut {NoStop}%
\bibitem [{\citenamefont {Jakli{\v c}}\ and\ \citenamefont {Prelov{\v
  s}ek}(2000)}]{JaP:AP00}%
  \BibitemOpen
  \bibfield  {author} {\bibinfo {author} {\bibfnamefont {J.}~\bibnamefont
  {Jakli{\v c}}}\ and\ \bibinfo {author} {\bibfnamefont {P.}~\bibnamefont
  {Prelov{\v s}ek}},\ }\href {http://dx.doi.org/10.1080/000187300243381}
  {\bibfield  {journal} {\bibinfo  {journal} {Adv. Phys.}\ }\textbf {\bibinfo
  {volume} {49}},\ \bibinfo {pages} {1} (\bibinfo {year} {2000})}\BibitemShut
  {NoStop}%
\bibitem [{\citenamefont {Manthe}\ and\ \citenamefont
  {Huarte-Larranaga}(2001)}]{MHL:CPL01}%
  \BibitemOpen
  \bibfield  {author} {\bibinfo {author} {\bibfnamefont {U.}~\bibnamefont
  {Manthe}}\ and\ \bibinfo {author} {\bibfnamefont {F.}~\bibnamefont
  {Huarte-Larranaga}},\ }\href {\doibase 10.1016/S0009-2614(01)01207-6}
  {\bibfield  {journal} {\bibinfo  {journal} {Chem. Phys. Lett.}\ }\textbf
  {\bibinfo {volume} {349}},\ \bibinfo {pages} {321 } (\bibinfo {year}
  {2001})}\BibitemShut {NoStop}%
\bibitem [{\citenamefont {Long}\ \emph {et~al.}(2003)\citenamefont {Long},
  \citenamefont {Prelov\ifmmode~\check{s}\else \v{s}\fi{}ek}, \citenamefont
  {El~Shawish}, \citenamefont {Karadamoglou},\ and\ \citenamefont
  {Zotos}}]{LPE:PRB03}%
  \BibitemOpen
  \bibfield  {author} {\bibinfo {author} {\bibfnamefont {M.~W.}\ \bibnamefont
  {Long}}, \bibinfo {author} {\bibfnamefont {P.}~\bibnamefont
  {Prelov\ifmmode~\check{s}\else \v{s}\fi{}ek}}, \bibinfo {author}
  {\bibfnamefont {S.}~\bibnamefont {El~Shawish}}, \bibinfo {author}
  {\bibfnamefont {J.}~\bibnamefont {Karadamoglou}}, \ and\ \bibinfo {author}
  {\bibfnamefont {X.}~\bibnamefont {Zotos}},\ }\href {\doibase
  10.1103/PhysRevB.68.235106} {\bibfield  {journal} {\bibinfo  {journal} {Phys.
  Rev. B}\ }\textbf {\bibinfo {volume} {68}},\ \bibinfo {pages} {235106}
  (\bibinfo {year} {2003})}\BibitemShut {NoStop}%
\bibitem [{\citenamefont {Aichhorn}\ \emph {et~al.}(2003)\citenamefont
  {Aichhorn}, \citenamefont {Daghofer}, \citenamefont {Evertz},\ and\
  \citenamefont {von~der Linden}}]{ADE:PRB03}%
  \BibitemOpen
  \bibfield  {author} {\bibinfo {author} {\bibfnamefont {M.}~\bibnamefont
  {Aichhorn}}, \bibinfo {author} {\bibfnamefont {M.}~\bibnamefont {Daghofer}},
  \bibinfo {author} {\bibfnamefont {H.~G.}\ \bibnamefont {Evertz}}, \ and\
  \bibinfo {author} {\bibfnamefont {W.}~\bibnamefont {von~der Linden}},\ }\href
  {http://dx.doi.org/10.1103/PhysRevB.67.161103} {\bibfield  {journal}
  {\bibinfo  {journal} {Phys. Rev. B}\ }\textbf {\bibinfo {volume} {67}},\
  \bibinfo {pages} {161103} (\bibinfo {year} {2003})}\BibitemShut {NoStop}%
\bibitem [{\citenamefont {Wei\ss{}e}\ \emph {et~al.}(2006)\citenamefont
  {Wei\ss{}e}, \citenamefont {Wellein}, \citenamefont {Alvermann},\ and\
  \citenamefont {Fehske}}]{WWA:RMP06}%
  \BibitemOpen
  \bibfield  {author} {\bibinfo {author} {\bibfnamefont {A.}~\bibnamefont
  {Wei\ss{}e}}, \bibinfo {author} {\bibfnamefont {G.}~\bibnamefont {Wellein}},
  \bibinfo {author} {\bibfnamefont {A.}~\bibnamefont {Alvermann}}, \ and\
  \bibinfo {author} {\bibfnamefont {H.}~\bibnamefont {Fehske}},\ }\href
  {\doibase 10.1103/RevModPhys.78.275} {\bibfield  {journal} {\bibinfo
  {journal} {Rev. Mod. Phys.}\ }\textbf {\bibinfo {volume} {78}},\ \bibinfo
  {pages} {275} (\bibinfo {year} {2006})}\BibitemShut {NoStop}%
\bibitem [{\citenamefont {Prelov\ifmmode~\check{s}\else \v{s}\fi{}ek}\ and\
  \citenamefont {Bon\ifmmode~\check{c}\else \v{c}\fi{}a}(2013)}]{PrB:SSSSS13}%
  \BibitemOpen
  \bibfield  {author} {\bibinfo {author} {\bibfnamefont {P.}~\bibnamefont
  {Prelov\ifmmode~\check{s}\else \v{s}\fi{}ek}}\ and\ \bibinfo {author}
  {\bibfnamefont {J.}~\bibnamefont {Bon\ifmmode~\check{c}\else \v{c}\fi{}a}},\
  }\enquote {\bibinfo {title} {Strongly correlated systems, numerical
  methods},}\ \ (\bibinfo  {publisher} {Springer},\ \bibinfo {address} {Berlin,
  Heidelberg},\ \bibinfo {year} {2013})\ Chap.\ \bibinfo {chapter} {Ground
  State and Finite Temperature Lanczos Methods}\BibitemShut {NoStop}%
\bibitem [{\citenamefont {Shannon}\ \emph {et~al.}(2004)\citenamefont
  {Shannon}, \citenamefont {Schmidt}, \citenamefont {Penc},\ and\ \citenamefont
  {Thalmeier}}]{SSP:EPJB04}%
  \BibitemOpen
  \bibfield  {author} {\bibinfo {author} {\bibfnamefont {N.}~\bibnamefont
  {Shannon}}, \bibinfo {author} {\bibfnamefont {B.}~\bibnamefont {Schmidt}},
  \bibinfo {author} {\bibfnamefont {K.}~\bibnamefont {Penc}}, \ and\ \bibinfo
  {author} {\bibfnamefont {P.}~\bibnamefont {Thalmeier}},\ }\href {\doibase
  10.1140/epjb/e2004-00156-3} {\bibfield  {journal} {\bibinfo  {journal} {Eur.
  Phys. J. B}\ }\textbf {\bibinfo {volume} {38}},\ \bibinfo {pages} {599}
  (\bibinfo {year} {2004})}\BibitemShut {NoStop}%
\bibitem [{\citenamefont {Zerec}\ \emph {et~al.}(2006)\citenamefont {Zerec},
  \citenamefont {Schmidt},\ and\ \citenamefont {Thalmeier}}]{ZST:PRB06}%
  \BibitemOpen
  \bibfield  {author} {\bibinfo {author} {\bibfnamefont {I.}~\bibnamefont
  {Zerec}}, \bibinfo {author} {\bibfnamefont {B.}~\bibnamefont {Schmidt}}, \
  and\ \bibinfo {author} {\bibfnamefont {P.}~\bibnamefont {Thalmeier}},\ }\href
  {\doibase 10.1103/PhysRevB.73.245108} {\bibfield  {journal} {\bibinfo
  {journal} {Phys. Rev. B}\ }\textbf {\bibinfo {volume} {73}},\ \bibinfo {eid}
  {245108} (\bibinfo {year} {2006})}\BibitemShut {NoStop}%
\bibitem [{\citenamefont {Schmidt}\ \emph {et~al.}(2007)\citenamefont
  {Schmidt}, \citenamefont {Thalmeier},\ and\ \citenamefont
  {Shannon}}]{STS:PRB07}%
  \BibitemOpen
  \bibfield  {author} {\bibinfo {author} {\bibfnamefont {B.}~\bibnamefont
  {Schmidt}}, \bibinfo {author} {\bibfnamefont {P.}~\bibnamefont {Thalmeier}},
  \ and\ \bibinfo {author} {\bibfnamefont {N.}~\bibnamefont {Shannon}},\ }\href
  {\doibase 10.1103/PhysRevB.76.125113} {\bibfield  {journal} {\bibinfo
  {journal} {Phys. Rev. B}\ }\textbf {\bibinfo {volume} {76}},\ \bibinfo
  {pages} {125113} (\bibinfo {year} {2007})}\BibitemShut {NoStop}%
\bibitem [{\citenamefont {Siahatgar}\ \emph {et~al.}(2012)\citenamefont
  {Siahatgar}, \citenamefont {Schmidt}, \citenamefont {Zwicknagl},\ and\
  \citenamefont {Thalmeier}}]{SSZ:NJP12}%
  \BibitemOpen
  \bibfield  {author} {\bibinfo {author} {\bibfnamefont {M.}~\bibnamefont
  {Siahatgar}}, \bibinfo {author} {\bibfnamefont {B.}~\bibnamefont {Schmidt}},
  \bibinfo {author} {\bibfnamefont {G.}~\bibnamefont {Zwicknagl}}, \ and\
  \bibinfo {author} {\bibfnamefont {P.}~\bibnamefont {Thalmeier}},\ }\href
  {http://dx.doi.org/10.1088/1367-2630/14/10/103005} {\bibfield  {journal}
  {\bibinfo  {journal} {New J. Phys.}\ }\textbf {\bibinfo {volume} {14}},\
  \bibinfo {pages} {103005} (\bibinfo {year} {2012})}\BibitemShut {NoStop}%
\bibitem [{\citenamefont {Schnack}\ and\ \citenamefont
  {Wendland}(2010)}]{ScW:EPJB10}%
  \BibitemOpen
  \bibfield  {author} {\bibinfo {author} {\bibfnamefont {J.}~\bibnamefont
  {Schnack}}\ and\ \bibinfo {author} {\bibfnamefont {O.}~\bibnamefont
  {Wendland}},\ }\href {http://dx.doi.org/10.1007/BF01609348} {\bibfield
  {journal} {\bibinfo  {journal} {Eur. Phys. J. B}\ }\textbf {\bibinfo {volume}
  {78}},\ \bibinfo {pages} {535} (\bibinfo {year} {2010})}\BibitemShut
  {NoStop}%
\bibitem [{\citenamefont {Schnack}\ and\ \citenamefont
  {Heesing}(2013)}]{ScH:EPJB13}%
  \BibitemOpen
  \bibfield  {author} {\bibinfo {author} {\bibfnamefont {J.}~\bibnamefont
  {Schnack}}\ and\ \bibinfo {author} {\bibfnamefont {C.}~\bibnamefont
  {Heesing}},\ }\href {http://dx.doi.org/10.1140/epjb/e2012-30546-7} {\bibfield
   {journal} {\bibinfo  {journal} {Eur. Phys. J. B}\ }\textbf {\bibinfo
  {volume} {86}},\ \bibinfo {pages} {46} (\bibinfo {year} {2013})}\BibitemShut
  {NoStop}%
\bibitem [{\citenamefont {Hanebaum}\ and\ \citenamefont
  {Schnack}(2014)}]{HaS:EPJB14}%
  \BibitemOpen
  \bibfield  {author} {\bibinfo {author} {\bibfnamefont {O.}~\bibnamefont
  {Hanebaum}}\ and\ \bibinfo {author} {\bibfnamefont {J.}~\bibnamefont
  {Schnack}},\ }\href {\doibase 10.1140/epjb/e2014-50360-5} {\bibfield
  {journal} {\bibinfo  {journal} {Eur. Phys. J. B}\ }\textbf {\bibinfo {volume}
  {87}},\ \bibinfo {eid} {194} (\bibinfo {year} {2014})}\BibitemShut {NoStop}%
\bibitem [{\citenamefont {Barra}\ \emph {et~al.}(1997)\citenamefont {Barra},
  \citenamefont {Gatteschi},\ and\ \citenamefont {Sessoli}}]{BGS:PRB97}%
  \BibitemOpen
  \bibfield  {author} {\bibinfo {author} {\bibfnamefont {A.~L.}\ \bibnamefont
  {Barra}}, \bibinfo {author} {\bibfnamefont {D.}~\bibnamefont {Gatteschi}}, \
  and\ \bibinfo {author} {\bibfnamefont {R.}~\bibnamefont {Sessoli}},\ }\href
  {\doibase 10.1103/PhysRevB.56.8192} {\bibfield  {journal} {\bibinfo
  {journal} {Phys. Rev. B}\ }\textbf {\bibinfo {volume} {56}},\ \bibinfo
  {pages} {8192} (\bibinfo {year} {1997})}\BibitemShut {NoStop}%
\bibitem [{\citenamefont {Glaser}\ \emph {et~al.}(2009)\citenamefont {Glaser},
  \citenamefont {Heidemeier}, \citenamefont {Krickemeyer}, \citenamefont
  {B{\"o}gge}, \citenamefont {Stammler}, \citenamefont {Fr{\"o}hlich},
  \citenamefont {Bill},\ and\ \citenamefont {Schnack}}]{GHK:IC09}%
  \BibitemOpen
  \bibfield  {author} {\bibinfo {author} {\bibfnamefont {T.}~\bibnamefont
  {Glaser}}, \bibinfo {author} {\bibfnamefont {M.}~\bibnamefont {Heidemeier}},
  \bibinfo {author} {\bibfnamefont {E.}~\bibnamefont {Krickemeyer}}, \bibinfo
  {author} {\bibfnamefont {H.}~\bibnamefont {B{\"o}gge}}, \bibinfo {author}
  {\bibfnamefont {A.}~\bibnamefont {Stammler}}, \bibinfo {author}
  {\bibfnamefont {R.}~\bibnamefont {Fr{\"o}hlich}}, \bibinfo {author}
  {\bibfnamefont {E.}~\bibnamefont {Bill}}, \ and\ \bibinfo {author}
  {\bibfnamefont {J.}~\bibnamefont {Schnack}},\ }\href
  {http://pubs.acs.org/doi/abs/10.1021/ic8016529} {\bibfield  {journal}
  {\bibinfo  {journal} {Inorg. Chem.}\ }\textbf {\bibinfo {volume} {48}},\
  \bibinfo {pages} {607} (\bibinfo {year} {2009})}\BibitemShut {NoStop}%
\bibitem [{\citenamefont {Glaser}\ \emph {et~al.}(2010)\citenamefont {Glaser},
  \citenamefont {Heidemeier}, \citenamefont {Theil}, \citenamefont {Stammler},
  \citenamefont {B{\"o}gge},\ and\ \citenamefont {Schnack}}]{GHT:DT10}%
  \BibitemOpen
  \bibfield  {author} {\bibinfo {author} {\bibfnamefont {T.}~\bibnamefont
  {Glaser}}, \bibinfo {author} {\bibfnamefont {M.}~\bibnamefont {Heidemeier}},
  \bibinfo {author} {\bibfnamefont {H.}~\bibnamefont {Theil}}, \bibinfo
  {author} {\bibfnamefont {A.}~\bibnamefont {Stammler}}, \bibinfo {author}
  {\bibfnamefont {H.}~\bibnamefont {B{\"o}gge}}, \ and\ \bibinfo {author}
  {\bibfnamefont {J.}~\bibnamefont {Schnack}},\ }\href
  {http://dx.doi.org/10.1039/b912593k} {\bibfield  {journal} {\bibinfo
  {journal} {Dalton Trans.}\ }\textbf {\bibinfo {volume} {39}},\ \bibinfo
  {pages} {192} (\bibinfo {year} {2010})}\BibitemShut {NoStop}%
\bibitem [{\citenamefont {Hoeke}\ \emph {et~al.}(2012)\citenamefont {Hoeke},
  \citenamefont {Gieb}, \citenamefont {M{\"u}ller}, \citenamefont {Ungur},
  \citenamefont {Chibotaru}, \citenamefont {Heidemeier}, \citenamefont
  {Krickemeyer}, \citenamefont {Stammler}, \citenamefont {B{\"o}gge},
  \citenamefont {Schr{\"o}der}, \citenamefont {Schnack},\ and\ \citenamefont
  {Glaser}}]{HGM:CS12}%
  \BibitemOpen
  \bibfield  {author} {\bibinfo {author} {\bibfnamefont {V.}~\bibnamefont
  {Hoeke}}, \bibinfo {author} {\bibfnamefont {K.}~\bibnamefont {Gieb}},
  \bibinfo {author} {\bibfnamefont {P.}~\bibnamefont {M{\"u}ller}}, \bibinfo
  {author} {\bibfnamefont {L.}~\bibnamefont {Ungur}}, \bibinfo {author}
  {\bibfnamefont {L.~F.}\ \bibnamefont {Chibotaru}}, \bibinfo {author}
  {\bibfnamefont {M.}~\bibnamefont {Heidemeier}}, \bibinfo {author}
  {\bibfnamefont {E.}~\bibnamefont {Krickemeyer}}, \bibinfo {author}
  {\bibfnamefont {A.}~\bibnamefont {Stammler}}, \bibinfo {author}
  {\bibfnamefont {H.}~\bibnamefont {B{\"o}gge}}, \bibinfo {author}
  {\bibfnamefont {C.}~\bibnamefont {Schr{\"o}der}}, \bibinfo {author}
  {\bibfnamefont {J.}~\bibnamefont {Schnack}}, \ and\ \bibinfo {author}
  {\bibfnamefont {T.}~\bibnamefont {Glaser}},\ }\href
  {http://dx.doi.org/10.1039/C2SC20649H} {\bibfield  {journal} {\bibinfo
  {journal} {Chem. Sci.}\ }\textbf {\bibinfo {volume} {3}},\ \bibinfo {pages}
  {2868} (\bibinfo {year} {2012})}\BibitemShut {NoStop}%
\bibitem [{\citenamefont {Farrell}\ \emph {et~al.}(2013)\citenamefont
  {Farrell}, \citenamefont {Coome}, \citenamefont {Probert}, \citenamefont
  {Goeta}, \citenamefont {Howard}, \citenamefont {Lemee-Cailleau},
  \citenamefont {Parsons},\ and\ \citenamefont {Murrie}}]{FCP:CEC13}%
  \BibitemOpen
  \bibfield  {author} {\bibinfo {author} {\bibfnamefont {A.~R.}\ \bibnamefont
  {Farrell}}, \bibinfo {author} {\bibfnamefont {J.~A.}\ \bibnamefont {Coome}},
  \bibinfo {author} {\bibfnamefont {M.~R.}\ \bibnamefont {Probert}}, \bibinfo
  {author} {\bibfnamefont {A.~E.}\ \bibnamefont {Goeta}}, \bibinfo {author}
  {\bibfnamefont {J.~A.~K.}\ \bibnamefont {Howard}}, \bibinfo {author}
  {\bibfnamefont {M.-H.}\ \bibnamefont {Lemee-Cailleau}}, \bibinfo {author}
  {\bibfnamefont {S.}~\bibnamefont {Parsons}}, \ and\ \bibinfo {author}
  {\bibfnamefont {M.}~\bibnamefont {Murrie}},\ }\href {\doibase
  10.1039/C3CE00042G} {\bibfield  {journal} {\bibinfo  {journal}
  {CrystEngComm}\ }\textbf {\bibinfo {volume} {15}},\ \bibinfo {pages} {3423}
  (\bibinfo {year} {2013})}\BibitemShut {NoStop}%
\bibitem [{\citenamefont {Barbara}\ \emph {et~al.}(1997)\citenamefont
  {Barbara}, \citenamefont {Gatteschi}, \citenamefont {Mukhin}, \citenamefont
  {Platonov}, \citenamefont {Popov}, \citenamefont {Tatsenko}, ,\ and\
  \citenamefont {Zvezdin}}]{BGM:1997}%
  \BibitemOpen
  \bibfield  {author} {\bibinfo {author} {\bibfnamefont {B.}~\bibnamefont
  {Barbara}}, \bibinfo {author} {\bibfnamefont {D.}~\bibnamefont {Gatteschi}},
  \bibinfo {author} {\bibfnamefont {A.~A.}\ \bibnamefont {Mukhin}}, \bibinfo
  {author} {\bibfnamefont {V.~V.}\ \bibnamefont {Platonov}}, \bibinfo {author}
  {\bibfnamefont {A.~I.}\ \bibnamefont {Popov}}, \bibinfo {author}
  {\bibfnamefont {A.~M.}\ \bibnamefont {Tatsenko}}, , \ and\ \bibinfo {author}
  {\bibfnamefont {A.~K.}\ \bibnamefont {Zvezdin}},\ }in\ \href@noop {} {\emph
  {\bibinfo {booktitle} {Proceedings of Seventh International Conference on
  Megagauss Magnetic Field Generation and Related Topics}}}\ (\bibinfo
  {address} {Sarov, 1996},\ \bibinfo {year} {1997})\ p.\ \bibinfo {pages}
  {853}\BibitemShut {NoStop}%
\bibitem [{\citenamefont {Regnault}\ \emph {et~al.}(2002)\citenamefont
  {Regnault}, \citenamefont {Jolic{\oe}ur}, \citenamefont {Sessoli},
  \citenamefont {Gatteschi},\ and\ \citenamefont {Verdaguer}}]{RJS:PRB02}%
  \BibitemOpen
  \bibfield  {author} {\bibinfo {author} {\bibfnamefont {N.}~\bibnamefont
  {Regnault}}, \bibinfo {author} {\bibfnamefont {T.}~\bibnamefont
  {Jolic{\oe}ur}}, \bibinfo {author} {\bibfnamefont {R.}~\bibnamefont
  {Sessoli}}, \bibinfo {author} {\bibfnamefont {D.}~\bibnamefont {Gatteschi}},
  \ and\ \bibinfo {author} {\bibfnamefont {M.}~\bibnamefont {Verdaguer}},\
  }\href {\doibase 10.1103/PhysRevB.66.054409} {\bibfield  {journal} {\bibinfo
  {journal} {Phys. Rev. B}\ }\textbf {\bibinfo {volume} {66}},\ \bibinfo
  {pages} {054409} (\bibinfo {year} {2002})}\BibitemShut {NoStop}%
\bibitem [{\citenamefont {Parois}(2010)}]{Par:PhD10}%
  \BibitemOpen
  \bibfield  {author} {\bibinfo {author} {\bibfnamefont {P.}~\bibnamefont
  {Parois}},\ }\emph {\bibinfo {title} {The effect of pressure on clusters,
  chains and single-molecule magnets}},\ \href@noop {} {Ph.D. thesis},\
  \bibinfo  {school} {Department of Chemistry, Faculty of Physical Sciences,
  University of Glasgow} (\bibinfo {year} {2010})\BibitemShut {NoStop}%
\bibitem [{\citenamefont {Schnack}(2009)}]{Sch:CMP09}%
  \BibitemOpen
  \bibfield  {author} {\bibinfo {author} {\bibfnamefont {J.}~\bibnamefont
  {Schnack}},\ }\href@noop {} {\bibfield  {journal} {\bibinfo  {journal}
  {Condens. Matter Phys.}\ }\textbf {\bibinfo {volume} {12}},\ \bibinfo {pages}
  {323} (\bibinfo {year} {2009})}\BibitemShut {NoStop}%
\bibitem [{\citenamefont {Glaser}\ and\ \citenamefont {Walleck}()}]{GlW:PC15}%
  \BibitemOpen
  \bibfield  {author} {\bibinfo {author} {\bibfnamefont {T.}~\bibnamefont
  {Glaser}}\ and\ \bibinfo {author} {\bibfnamefont {S.}~\bibnamefont
  {Walleck}},\ }\href@noop {} {}\bibinfo {note} {Private
  communication}\BibitemShut {NoStop}%
\bibitem [{\citenamefont {Novak}\ \emph {et~al.}(1995)\citenamefont {Novak},
  \citenamefont {Sessoli}, \citenamefont {Caneschi},\ and\ \citenamefont
  {Gatteschi}}]{NSC:JMMM95}%
  \BibitemOpen
  \bibfield  {author} {\bibinfo {author} {\bibfnamefont {M.~A.}\ \bibnamefont
  {Novak}}, \bibinfo {author} {\bibfnamefont {R.}~\bibnamefont {Sessoli}},
  \bibinfo {author} {\bibfnamefont {A.}~\bibnamefont {Caneschi}}, \ and\
  \bibinfo {author} {\bibfnamefont {D.}~\bibnamefont {Gatteschi}},\ }\href@noop
  {} {\bibfield  {journal} {\bibinfo  {journal} {J. Magn. Magn. Mater.}\
  }\textbf {\bibinfo {volume} {146}},\ \bibinfo {pages} {211} (\bibinfo {year}
  {1995})}\BibitemShut {NoStop}%
\bibitem [{\citenamefont {Zvezdin}\ \emph {et~al.}(1998)\citenamefont
  {Zvezdin}, \citenamefont {Lubashevskii}, \citenamefont {Levitin},
  \citenamefont {Platonov},\ and\ \citenamefont {Tatsenko}}]{ZLL:PU98}%
  \BibitemOpen
  \bibfield  {author} {\bibinfo {author} {\bibfnamefont {A.~K.}\ \bibnamefont
  {Zvezdin}}, \bibinfo {author} {\bibfnamefont {I.~A.}\ \bibnamefont
  {Lubashevskii}}, \bibinfo {author} {\bibfnamefont {R.~Z.}\ \bibnamefont
  {Levitin}}, \bibinfo {author} {\bibfnamefont {V.~V.}\ \bibnamefont
  {Platonov}}, \ and\ \bibinfo {author} {\bibfnamefont {O.~M.}\ \bibnamefont
  {Tatsenko}},\ }\href {http://stacks.iop.org/1063-7869/41/i=10/a=A06}
  {\bibfield  {journal} {\bibinfo  {journal} {Physics-Uspekhi}\ }\textbf
  {\bibinfo {volume} {41}},\ \bibinfo {pages} {1037} (\bibinfo {year}
  {1998})}\BibitemShut {NoStop}%
\bibitem [{\citenamefont {Fern\'andez}\ \emph {et~al.}(1998)\citenamefont
  {Fern\'andez}, \citenamefont {Luis},\ and\ \citenamefont
  {Bartolom\'e}}]{FLB:PRL98}%
  \BibitemOpen
  \bibfield  {author} {\bibinfo {author} {\bibfnamefont {J.~F.}\ \bibnamefont
  {Fern\'andez}}, \bibinfo {author} {\bibfnamefont {F.}~\bibnamefont {Luis}}, \
  and\ \bibinfo {author} {\bibfnamefont {J.}~\bibnamefont {Bartolom\'e}},\
  }\href {\doibase 10.1103/PhysRevLett.80.5659} {\bibfield  {journal} {\bibinfo
   {journal} {Phys. Rev. Lett.}\ }\textbf {\bibinfo {volume} {80}},\ \bibinfo
  {pages} {5659} (\bibinfo {year} {1998})}\BibitemShut {NoStop}%
\bibitem [{\citenamefont {Luis}\ \emph {et~al.}(2000)\citenamefont {Luis},
  \citenamefont {Mettes}, \citenamefont {Tejada}, \citenamefont {Gatteschi},\
  and\ \citenamefont {de~Jongh}}]{LMT:PRL00}%
  \BibitemOpen
  \bibfield  {author} {\bibinfo {author} {\bibfnamefont {F.}~\bibnamefont
  {Luis}}, \bibinfo {author} {\bibfnamefont {F.~L.}\ \bibnamefont {Mettes}},
  \bibinfo {author} {\bibfnamefont {J.}~\bibnamefont {Tejada}}, \bibinfo
  {author} {\bibfnamefont {D.}~\bibnamefont {Gatteschi}}, \ and\ \bibinfo
  {author} {\bibfnamefont {L.~J.}\ \bibnamefont {de~Jongh}},\ }\href {\doibase
  10.1103/PhysRevLett.85.4377} {\bibfield  {journal} {\bibinfo  {journal}
  {Phys. Rev. Lett.}\ }\textbf {\bibinfo {volume} {85}},\ \bibinfo {pages}
  {4377} (\bibinfo {year} {2000})}\BibitemShut {NoStop}%
\end{thebibliography}

\end{document}